\documentclass[11pt,a4paper]{article}
\usepackage{amsmath,amsfonts,amssymb,amsthm}
\usepackage[mathscr]{eucal}
\newcommand\cA{{\mathcal A}}

\newcommand\cD{{\mathcal D}} 
 
\newcommand\cF{{\mathcal F}}

\newcommand\cI{{\mathcal I}} 
\newcommand\cJ{{\mathcal J}} 
\newcommand\cK{{\mathcal K}}

\newcommand\cN{{\mathcal N}}

\newcommand\mvector{\boldsymbol}

\newcommand\vc{\mvector{c}}
\newcommand\vd{\mvector{d}}
\newcommand\ve{\mvector{e}}
\newcommand\vf{\mvector{f}}

\newcommand\vk{\mvector{k}}

\newcommand\vp{\mvector{p}}
\newcommand\vq{\mvector{q}}

\newcommand\vu{\mvector{u}}
\newcommand\vv{\mvector{v}}

\newcommand\vx{\mvector{x}}
\newcommand\vy{\mvector{y}}

\newcommand\vF{\mvector{F}}
\newcommand\vG{\mvector{G}}

\newcommand\vP{\mvector{P}}
\newcommand\vQ{\mvector{Q}}

\newcommand\vxi{\mvector{\xi}}

\newcommand\vlambda{\mvector{\lambda}}
\newcommand\vLambda{\mvector{\Lambda}}

\newcommand\vvarphi{\mvector{\varphi}}

\newcommand\vzero{\mvector{0}}

\newcommand\field{\mathbb}

\newcommand\R{\field{R}}

\newcommand\C{\field{C}}
\newcommand\Z{\field{Z}}

\newcommand\N{\field{N}}

\newcommand\K{\field{K}}

\newcommand\bbP{\mathbb{P}}

\newcommand\Div{\operatorname{div}}

\newcommand\res{\operatorname{res}}

\newcommand\rmd{\mathrm{d}\mspace{1mu}}

\newcommand\CPOne{\ensuremath{\C\mathbb{P}^1}}
\newcommand\CP{\ensuremath{\C\bbP}}
\newcommand\rmi{\mathrm{i}\mspace{1mu}}

\newcommand\Dt{\frac{\mathrm{d}\phantom{t} }{\mathrm{d}\mspace{1mu} t}}

\newcommand\pder[2]{\dfrac{\partial #1 }{\partial #2}} 
\newcommand\abs[1]{\lvert #1 \rvert}
\newcommand\norm[1]{\lVert #1 \rVert}

\newcommand\mtext[1]{\quad\text{#1}\quad}

\newcommand\defset[2]{\left\{{#1}\;\vert \;\; {#2} \,\right\}}

\usepackage[square]{natbib}
\theoremstyle{plain}
\newtheorem{theorem}{Theorem}
\newtheorem{lemma}{Lemma}
\newtheorem{proposition}{Proposition}
\newtheorem{corollary}{Corollary}

\newtheorem{conjecture}{Conjecture}
\newtheoremstyle{note}{\topsep}{\topsep}{\slshape}{}{\scshape}{}{ }{}
\theoremstyle{note}

\newtheorem{remark}{Remark}
\theoremstyle{definition}
\newtheorem{definition}{Definition}
\numberwithin{equation}{section}
\numberwithin{theorem}{section}
\numberwithin{lemma}{section}
\numberwithin{proposition}{section}
\numberwithin{corollary}{section}
\numberwithin{conjecture}{section}
\numberwithin{remark}{section}
\begin{document}
\thispagestyle{empty}
{\flushright{\textsf{submitted to Journal of Mathematical Physics}}}
\vspace*{1em}
\begin{center}
\LARGE\textbf{Differential Galois obstructions for integrability of homogeneous
Newton equations}
\end{center}
\vspace*{0.5em}
\begin{center}
\large  Maria Przybylska
\end{center}
\vspace{2em}
\hspace*{2em}\begin{minipage}{0.8\textwidth}
\small
 Institut Fourier, UMR 5582 du CNRS,
Universit\'e de Grenoble I, \\\quad
100 rue des Maths,
BP 74, 38402 Saint-Martin d'H\`eres Cedex, France, and \\[0.5em]
Toru\'n Centre for Astronomy,
  N.~Copernicus University, 
  Gagarina 11,\\\quad PL-87--100 Toru\'n, Poland,
  (e-mail: Maria.Przybylska@astri.uni.torun.pl)
\end{minipage}\\[2.5em]
\vbox to 1.5em {}
{\small \textbf{Abstract.}  
In this paper we formulate  necessary conditions for the integrability in the
Jacobi sense of Newton equations $\ddot \vq=-\vF(\vq)$, where $\vq\in\C^n$ and
all components of $\vF$ are polynomial and homogeneous of the same degree $l$. These
conditions are derived from an analysis of the differential Galois group of the
variational equations along special particular solutions of the Newton
equations.
 We show that, taking
all admissible particular solutions, we restrict considerably the set of
Newton's
equations
satisfying the necessary conditions for the integrability. Moreover, we apply the obtained conditions for a detailed analysis of the Newton
equations  with two degrees of freedom (i.e., $n=2$). We demonstrate the
strength of the obtained results analyzing general cases with $\deg F_i =l< 4$.  For $l=3$ we found an integrable case  when  the Newton equations have two polynomial first integrals and both of them are of degree four in the momenta $p_1=\dot q_1$, and $p_2=\dot q_2$. 
Moreover, for an arbitrary $k$, we found a family of Newton equations depending on one parameter $\lambda$. For an arbitrary value of $\lambda$ one quadratic in the momenta first integral exist. 
We distinguish infinitely many values of $\lambda$ for which the system 
is integrable or superintegrable with additional polynomial first integrals which seemingly can be of  an
arbitrarily high degree with respect to the momenta.
}

\section{Introduction}

In this paper we study the integrability  of the following class of Newton's
equations
\begin{equation}
\label{eq:newton} 
\ddot \vq= -\vF(\vq), \qquad\vq=(q_1,\ldots,q_n)^T\in\C^n.
\end{equation}
 In spite of the fact that such equations already 
appeared at the beginning of the mechanics in the Newton Principia
\cite{Newton:1687::}, there is not too many rigorous results concerning their 
integrability. The only exception is the case of potential forces, i.e., the
case when 
\begin{equation*}
 \vF(\vq)=\pder{V}{\vq}(\vq),
\end{equation*}
for a certain function $V$. In fact, for potential forces
equations~\eqref{eq:newton} are equivalent to Hamilton's equations
\begin{equation}
 \dot \vq = \vp, \qquad\dot \vp = -\pder{V}{\vq}(\vq),
\end{equation} 
with natural Hamiltonian
\begin{equation}
\label{eq:nham}
 H = \frac{1}{2}\vp^T\vp +V(\vq). 
\end{equation} 
First of all, for Hamiltonian systems we have a precise notion of the
integrability i.e. the integrability in the Liouville sense. Moreover, for such
systems we have powerfull methods and tools for integrability studies.  
This is why in this paper  we investigate  the   integrability problem  for
Newton's
equations with non-potential forces.  However, for such equations
`integrability'
is not well defined, so  we should first specify what, in the considered
context,
the 
integrability means.

The  oldest notion of the integrability is related to quadratures. We say
that  the considered system of differential equations is integrable by
quadratures
if all its solutions can be obtained  in a closed form by a finite number of
algebraic operations
(including inversion of functions) and quadratures, i.e.   integrals
of `known' functions, see e.g. \cite{Arnold:88::}. In fact, already in his
Principia Newton  reduced to quadratures the simplest problems of dynamics (see
theorems XXXIX, XLI, LIII and so on, see also comments in
\cite{Gallavotti:83::}). A similar definition of integrability was
applied by Poincar\'e  \cite{Poincare:1892::}.

The integrability by quadratures was a prototype for  other definitions of
integrability. As a matter of fact, many of these new definitions can be
considered as 
necessary conditions for the integrability by quadratures.   Generally,  if
for a given dynamical system 
\begin{equation}
\label{eq:dsg}
\dot \vx =\vv (\vx),\qquad \vx=(x_1,\ldots x_n)^T\in\C^n,
\end{equation}
there  exists  a sufficient number of tensor invariants, i.e. tensor fields that
are constant along phase curves, then the system is solvable by quadratures. 
The most known two examples are the following.
 If system~\eqref{eq:dsg} possesses $n-1$ functionally independent first
integrals, then it is integrable by quadratures. 
The Lie theorem says that if system~\eqref{eq:dsg} has  $n$ symmeties, i.e., if there exist vector fields  $\vu_1(\vx)=\vv(\vx), \vu_2(\vx),\ldots, \vu_n(\vx)$ which are
linearly independent and $[\vu_i,\vu_j]=\vzero$, for $i,j=1,\ldots,n$, then
system~\eqref{eq:dsg} is integrable by quadratures. 

In \cite{Jacobi:1884::} C.~G.~J.~Jacobi developed, introduced by L.~Euler, a
method of the integrable factor. The Jacobi Last Multiplier of
equation~\eqref{eq:dsg} is an invariant  $n$-form
\begin{equation*}
 \mu=h(\vx)\mathrm{d}x_1\wedge\cdots\wedge\mathrm{d}x_n.
\end{equation*}
Invariance of $\mu$ means that
\[
\Div (h\vv):=\sum_{i=1}^n\frac{\partial \phantom{-}}{\partial x_i}(h v_i)=0.
\]
\begin{theorem}[Jacobi]
Assume that system~\eqref{eq:dsg} possesses $n-2$ functionally independent first
integrals and a Last Jacobi Multiplier.  Then it is integrable by quadratures. 
\end{theorem}
We can also recall  the well known fact that the integrability of a  Hamiltonian
system in the Liouville sense implies its integrability by quadratures, see e.g.
\cite{Arnold:78::}. For more details about the integrability by quadratures see
\cite{Arnold:88::,Fedorov:06::}.

For the purpose of this paper we introduce the following definition. 
\begin{definition}
 We say that system~\eqref{eq:dsg} is integrable in the Jacobi sense if it
admits an invariant $n$-form and possesses $n-2$ functionally independent first
integrals.
\end{definition}
This definition of the integrability is very often applied for nonholonomic
systems, 
see e.g.
\cite{Arnold:88::,Fedorov:06::,Kozlov:96::,Chaplygin:81::,Jovanovic:2001::}.

Let us notice that Newton's equations~\eqref{eq:newton} can be written in the
following form
\begin{equation}
 \label{eq:newton1}
\dot\vq=\vp, \qquad \dot\vp = -\vF(\vq),
\end{equation} 
so they admit 
\begin{equation*}
 \mu = \rmd q_1\wedge \cdots \wedge\rmd q_n \wedge \rmd p_1\wedge \cdots
\wedge\rmd p_n,
\end{equation*}
as an invariant $2n$-form. Thus, in a case when equations~\eqref{eq:newton1} are
not Hamiltonian, we say that they are integrable if they are integrable in the
Jacobi sense. 

The integrability does not depend on the chosen coordinates, however, the form
of
equations~\eqref{eq:newton1} and the homogeneity of $\vF$  do.  
A linear change of variables  given by  $\vq=A\vQ$, $\vp=A\vP$, where
$A\in \mathrm{GL}(n,\C)$, transforms equations~\eqref{eq:newton1} into
\begin{equation}
 \label{eq:newton1PQ}
\dot\vQ=\vP, \qquad \dot\vP = -\vF_A(\vQ),\mtext{where} \vF_A(\vQ):=
A^{-1}\vF(A\vQ). 
\end{equation}  
Hence it is reasonable to  divide all forces into disjoint equivalence classes.
We say that forces $\vF$ and $\vG$ are equivalent iff there exists $A\in
\mathrm{GL}(n,\C)$ such that $\vG=\vF_A$. 
Later, specifying a force we just give a
representative of equivalent forces in the above sense. 

Our main purpose is to consider  Newton's equations which are not Hamiltonian.  
Having this in mind, we introduce the following definition.
\begin{definition}
 \label{def:pot}
We say that a force $\vF$ is potential iff there exists $A\in
\mathrm{GL}(n,\C)$ and $V\in\C[\vq]$ such that $\vF_A=\nabla V$.
\end{definition}

For Hamiltonian systems there are many approaches to the integrability problem:
the Hamilton-Jacobi theory, perturbation techniques,
normal forms, splitting of separatrices and, recently, the Morales-Ramis theory
based
on differential Galois theory. Many integrable Hamiltonian
systems were found and exhaustively analyzed by means of  these methods. 
For non-Hamiltonian systems  we
have not so many various approaches.

The main aim of this paper is to formulate  necessary conditions of
integrability in the Jacobi sense for Newton systems \eqref{eq:newton1}.  
In this aim, we use an approach which is based on an
investigation of variational equations for a particular complexified solution.
This approach was developed by  
 by S.L.~Ziglin \cite{Ziglin:82::b,Ziglin:83::b} for Hamiltonian systems.
However, the basic idea of his method can be used for a study of  the
integrability
of an arbitrary dynamical system.   
The fact that the considered system possesses  a meromorphic first integral
implies that  there exists a rational function which is an invariant of the 
monodromy group of the variational equations. This gives a restriction on the
monodromy group.  For the first time this method  was applied for proving the
non-integrability of a non-Hamiltonian system by S.L.~Ziglin in
\cite{Ziglin:96::,Ziglin:98::,MR1748212,MR2021137}.

The Ziglin theory appeared to be a very strong tool for proving the
non-integrability of Hamiltonian systems. Thanks to works of A.~Baider,
R.~.C.~Churchill, J.~J.~Morales, J.-P.~Ramis, D.~L.~Rod, C.~Sim\'o and
M.~F.~Singer  the Ziglin theory was considerably developed,
see~\cite{Churchill:96::b,Morales:99::c,Morales:99::b,Morales:01::b1} and
references
therein.  The main idea of this development  is following.  As in the Ziglin
theory, we investigate the variational equations along a non-equilibrium
solution. But, instead of the monodromy group of the variational equations, we
investigate their differential Galois group.  This change gives the substantial
profit because it is much easier to investigate the differential Galois group of
the
given equations than their monodromy group, and the differential Galois group
contains 
the monodromy group as a subgroup.   The main theorem due to Morales
and Ramis states that if a Hamiltonian systems is integrable in the Liouville
sense, then the identity component of the differential Galois group of
variational equations along a non-equilibrium solution is Abelian. Some parts of
the described differential Galois approach to the integrability can be adopted
to non-Hamiltonian systems. The key implication is the following. If a system
possesses a
meromorphic first integral, then the differential Galois group of variational
equations along a non-equilibrium solution has a rational invariant.  In
\cite{Maciejewski:02::f} the reader finds the first application of the
differential Galois approach for proving non-integrability of a non-Hamiltonian
system.  This type of the   integrability analysis was also applied for a
nonholonomic system called 
the Suslov system in
\cite{mp:04::b}.  In this paper we apply the differential Galois approach
for integrability
studies of Newton's equations. 

To formulate the main results of this paper we have to fix our assumptions.
Hence, if it is not otherwise stated, we assume that force
$\vF=(F_1,\ldots,F_n)$ is polynomial and homogeneous, i.e., $F_i\in\C[\vq]$, and
$\deg F_i=l>1$ for $i=1,\ldots, n$. Moreover, we always
write  $\deg F_i=l=k-1$, and letter $k$ is used only in this context (if $\vF$
is a potential force, then $k$ is the degree of the potential). Thus we have
always $k>2$.
By a direction  in $\C^n$ we understand an equivalence class of parallel
non-zero vectors. Of course a direction is determined just by one non-zero
vector.
\begin{definition}
  A direction $\vd\in\C^n$ is a Darboux 
point of a force $\vF(\vq)$, iff $\vd$ is parallel to $\vF(\vd)$, i.e., $\vd
\wedge \vF(\vd)=\vzero$,   and $\vF(\vd)\neq\vzero$.
\end{definition}
To find all Darboux points of a given force it is enough to find all non-zero
solutions of equations
\begin{equation}
 \vF(\vd) = \vd,
\end{equation} 
and take those which give different directions. To simplify exposition, we
always use such normalization  that a Darboux point $\vd$ of $\vF$ satisfies
the above equation. The set of all Darboux points of a given force $\vF$ is
denoted by $\cD_{\vF}$. It can be an empty, infinite or finite set. If
$\cD_{\vF}$ is
finite,  then for a generic $\vF$ it has $D(n,k):=[(k-1)^n-1]/(k-2)$ elements.

A Darboux point $\vd$
defines a particular solution of system~\eqref{eq:newton1}  by
\begin{equation}
\label{eq:part}
  \vq(t) = \varphi(t) \vd, \qquad \vp(t)=\dot \varphi(t)\vd,
\end{equation}
where $\varphi(t)$ satisfies the equation
\begin{equation*}
\ddot\varphi = -\varphi^{k-1}.
\end{equation*}
This equation determines a family of hyperelliptic curves
\begin{equation}
\label{eq:hypel}
{\dot\varphi}^2 = \frac{2}{k}\left(\varepsilon -\varphi^{k}\right), 
\end{equation}  
depending on a parameter $\varepsilon\in\C^\star$.

The variational equations along solution~\eqref{eq:part} have the form
\begin{equation}
  \label{eq:var}
   \dot \vx = \vy, \qquad
  \dot \vy =-\varphi(t)^{k-2}\vF'(\vd) \vx,
\end{equation}
or simply
\begin{equation}
 \label{eq:varso}
 \ddot \vx =-\varphi(t)^{k-2}\vF'(\vd) \vx,
\end{equation} 
where $\vF'(\vd)$ is the Jacobi matrix of $\vF$ calculated at a Darboux point
$\vd$. Let
us
assume that this matrix is diagonalizable.  Then,  in an appropriate basis
equations \eqref{eq:varso} have the form 
\begin{equation}
  \label{eq:unve}
  \ddot \eta_i = -\lambda_i \varphi(t)^{k-2}\eta_i,
\qquad i=1,\ldots, n,
\end{equation}
where $\lambda_1, \ldots, \lambda_n$  are
eigenvalues
of $\vF'(\vd)$. By homogeneity of $\vF$ one of the eigenvalues, let us say
${\lambda}_n$, is
$(k-2)$. We denote the rest of eigenvalues  by 
$\vlambda=\vlambda(\vd)=(\lambda_1,\ldots,\lambda_{n-1})\in\C^{n-1}$.

Our first theorem is the following.
\begin{theorem}
 \label{thm:mu}
Assume that the Newton system~\eqref{eq:newton} with polynomial homogeneous
right-hand sides of degree greater than one  is integrable in the Jacobi sense.
If $\vd$ is a Darboux point of $\vF$, such that $\vF'(\vd)$ is semi-simple, then
the
identity component of the differential Galois group of variational equations
along the particular solution defined by $\vd$ is Abelian. 
\end{theorem}
Our next theorem gives a computable criterion for the integrability in the
Jacobi sense. 
\begin{theorem}
\label{thm:MoRa}
Assume that the Newton system~\eqref{eq:newton} with polynomial homogeneous
right-hand sides of degree $l=k-1 >1$ is integrable in the Jacobi sense. If
$\vd$ is a Darboux point of $\vF$,  such that $\vF'(\vd)$ is semi-simple,  and
$\lambda_1, \ldots, \lambda_n$ are  eigenvalues of  $\vF'(\vd)$,  then 
$(k,\lambda_i)$ for $i=1,\ldots,n$ belong to the
following list 
\begin{equation}
 \label{eq:tab}
\begin{array}{crcr}
\text{case} &  k & \lambda  &\\[0.5em]
\hline
\vbox to 1.8em{} 1.  & k & p + \dfrac{k}{2}p(p-1) & \\[0.9em]
2. & k & \dfrac 1 {2}\left(\dfrac {k-1} {k}+p(p+1)k\right)  & \\[0.9em]
3. & 3 &  -\dfrac 1 {24}+\dfrac 1 {6}\left( 1 +3p\right)^2, & -\dfrac 1
{24}+\dfrac 3 {32}\left(  1  +4p\right)^2 \\[0.7em]
 & & -\dfrac 1 {24}+\dfrac 3 {50}\left(  1  +5p\right)^2,  &
-\dfrac1{24}+\dfrac{3}{50}\left(2 +5p\right)^2 \\[0.9em]
4. & 4 & -\dfrac 1 8 +\dfrac{2}{9} \left( 1+ 3p\right)^2 & \\[0.9em]
5. & 5 & -\dfrac 9 {40}+\dfrac 5 {18}\left(1+ 3p\right)^2, & -\dfrac 9
{40}+\dfrac 1 {10}\left(2
  +5p\right)^2  \\[1em]
\hline
\end{array}
\end{equation}
where $p$ is an integer.  Moreover, among $\lambda_1,\dots \lambda_n$ at most
two belong to the first 
item  of the above table.
\end{theorem}
The above theorem follows from Theorem~\ref{thm:mu} and the fact that each of
the second order variational equations in~\eqref{eq:unve} can be transformed
into
a Gauss hypergeometric equation for which the differential Galois group is
known.

Our next result is of a different nature. It is easy to understand that an
application of Theorem~\ref{thm:MoRa} to a general multi-parameter family of $\vF$ gives
rather an unsatisfactory result: we distinguish infinitely many families of
$\vF$
that satisfy conditions of~Theorem~\ref{thm:MoRa}. One can try to find more
than one Darboux point and apply~Theorem~\ref{thm:MoRa} to all of them. But in
practice, without any guiding idea, an analysis of the obtained conditions is at
least difficult.

To describe our approach, let us define for a Darboux point $\vd$ of $\vF$
quantities  $\Lambda_i=\lambda_i-1$, for $i=1,\ldots, n-1$, where $\lambda_i$
are eigenvalues of $\vF'(\vd)$. We denote them by
$\vLambda=\vLambda(\vd)=(\Lambda_1,\ldots,\Lambda_{n-1})\in\C^{n-1}$. Let
$\tau_i$, for $i=0,\ldots,n-1$, denote elementary symmetric polynomials in
$(n-1)$ variables.
The following theorem shows that for a given $\vF$, quantities $\vLambda(\vd)$
calculated at different Darboux points are not arbitrary. It gives $n$ relations
among $\vLambda(\vd)$ calculated over all Darboux points $\vd\in\cD_{\vF}$.
\begin{theorem}
 \label{thm:kojot}
Assume that $\vF$ has exactly $D(n,k)$  Darboux points $\vd\in\cD_{\vF}$. Then
$\vLambda(\vd)$ satisfy the following relations:
\begin{equation}
 \label{eq:rkoj}
 \sum_{\vd\in\mathcal{D}_{\vF}}\frac{ \tau_1(\vLambda(\vd))^r
 }{\tau_{n-1}(\vLambda(\vd))}=(-1)^{n}(n+k-2)^r,
\end{equation} 
 or, alternatively
\begin{equation}
 \label{eq:rksym}
 \sum_{\vd\in\mathcal{D}_{\vF}}\frac{ \tau_r(\vLambda(\vd))
 }{\tau_{n-1}(\vLambda(\vd))} =
(-1)^{n-r-1}\sum_{i=0}^{r}\binom{n-i-1}{r-i}(k-1)^{i},
\end{equation} 
 for $0\leq r\leq n-1$.
\end{theorem}
The most important consequence of this theorem is the following. We show
that
for a fixed $k$, the set of solutions of \eqref{eq:rkoj} or \eqref{eq:rksym},
such that for each $\vd\in\cD_{\vF}$  all components of $\vlambda(\vd)$ satisfy
conditions of Theorem~\ref{thm:MoRa}  is finite. Moreover, there is an
algorithmic way to find all of those solutions. In other words, the above
theorem gives a possibility to investigate the integrability of a general polynomial Newton
system with a fixed degree of homogeneity.

It is worth mentioning here that the main idea which allows to derive the result
contained in Theorem~\ref{thm:kojot} is to consider the following auxiliary
system
\begin{equation}
 \label{eqaux}
\Dt \vq = \vF(\vq)
\end{equation}
associated with Newton's equations~\eqref{eq:newton1}.  It is a homogeneous
first order system, so we can perform the Kovalevskaya analysis for it, see e.g.
\cite{Kozlov:96::}.
Then it appears that the introduced quantities $\Lambda_i$ are the Kovalevskaya
exponents calculated at Darboux points of system~\eqref{eqaux}. For more
details, see Section~\ref{sec:glo}.

Without doubt the integrability is not a generic phenomenon. Thus one would like
to
weaken the assumptions of Theorem~\ref{thm:kojot} and consider forces $\vF$ with
a
smaller number of Darboux points. We do not know how to treat such cases for
an arbitrary $n$, but for $n=2$ we present an approach which allows to formulate
a
theorem much stronger than Theorem~\ref{thm:kojot}, for details see  Section~\ref{sec:appgen}.

Applications of our  general theorems  for Newton's equations
with
two degrees of freedom enabled us to distinguish for an
arbitrary $k$ a one parameter family  for which  the necessary integrability
conditions formulated in Theorem~\ref{thm:MoRa}  are
seemingly  sufficient. An analysis given in Section~\ref{ssec:intfam} supports
the following conjecture.
\begin{conjecture}
 \label{con:ifam}
Assume that $(k,\lambda)$ belongs to an item of table~\eqref{eq:tab}. Then
Newton equations
\begin{equation}
\label{eq:nifam}
 \dot q_1 = p_1, \quad \dot p_1=-\lambda q_1 q_2^{k-2}\qquad \dot q_2 = p_2,
\quad \dot p_2 =q_2^{k-1} 
\end{equation}
 are integrable in the Jacobi sense with polynomial first integrals $I_1$ and
$I_2$, where 
\begin{equation*}
 I_1=\frac{1}{2}p_2^2 +\frac{1}{k}q_2^k.
\end{equation*}
Moreover, for an arbitrary $M>0$
we find $\lambda$ such that the degree of $I_2$ with respect to momenta is
greater than $M$ and there is no an additional polynomial  first integral
independent with $I_1$ and degree with respect momenta smaller or equal $M$. 
Additionally, if $(k,\lambda)$ belongs to an item different from 1 in
table~\eqref{eq:tab}, then there exist two additional polynomial first integrals $I_2$
and $I_3$ which are functionally independent together with $I_1$.
\end{conjecture}
The plan of this paper is the following. In Section~\ref{sec:gener} definitions
and some general results concerning the relations between first integrals of
dynamical systems and invariants of their differential Galois group are
presented.  In Section~\ref{sec:part} particular solutions and
variational equations are calculated. Section~\ref{sec:nec} is crucial because
contains necessary conditions of the  Jacobi integrability. In
Section~\ref{sec:glo}  consequences
 of the existence of more particular solutions related to Darboux points for the
integrability analysis are 
considered. Section~\ref{sec:appgen} is a compendium of all general results for
systems of two 
Newton's equations. In the next section we apply these results to the general
classes of  two Newton's 
homogeneous equations  with the right-hand sides of degree two and three and for
such systems the 
integrability analysis is almost complete.
Some partial general results for two Newton equations with higher degrees of homogeneity
are also presented.
 Appendix A contains two approaches to the problem how to recognized that the
analyzed Newton's 
system is Hamiltonian. In Appendix B the basic facts about higher order
variational equations and their 
applications are shown. The necessary integrability conditions expressed by
means  of higher order 
variational equations appeared to be a very effective tool for the analysis of Hamilton equations. 
In this part of the paper we formulate a conjecture that they can be also
applied for the Newton equations and 
examples show the validity of this hypothesis.

\section{Basic facts from general theory} 
\label{sec:gener}
Let us consider a system of differential equations
\begin{equation}
\label{eq:ds}
\Dt \vx = \vv(\vx), \qquad \vx\in U \subset \C^n, \quad t\in\C,
\end{equation} 
where $\vv$ is a holomorphic vector field in the considered domain $U$.  For a
non-equilibrium  particular solution $\vvarphi(t)$ we consider also the
variational equations along  $\vvarphi(t)$, i.e.,
\begin{equation}
\label{eq:vds}
\Dt\vxi=A(t)\vxi, \qquad A(t)=\pder{\vv}{\vx}(\vvarphi(t)).
\end{equation}
The considered particular solution $\vvarphi(t)$ defines a Riemann surface
$\Gamma$  immersed  in $ \C^n$ with $t$ as a local parameter.  The right hand
sides of variational
equations~\eqref{eq:vds} depend in fact on a point $\Gamma$.
 
 If $F:\C^n \supset W \rightarrow \C$ is a holomorphic function defined in a
certain connected neighborhood  $W$ of the solution $\varphi(t)$, then, by
definition,  the leading term $f$ of $F$  is the lowest order  non-vanishing term   of
expansion
 \begin{equation}
F(\vvarphi(t) +\vxi)= F_m(\vxi) + O(\norm{\vxi}^{m+1}),  \qquad F_m\neq 0, 
\end{equation}
i.e., $f(\vxi):=F_m(\vxi)$. 
Note that  $f(\vxi)$ is a homogeneous polynomial of variables
$\vxi=(\xi_1,\ldots, \xi_n)$ of degree $m$; its coefficients are polynomials in
$\vvarphi(t)$.  If $F$ is a meromorphic function,
then $F=P/Q$  for certain holomorphic functions $P$ and $Q$. In this case, the
leading term  $f$ of $F$  is defined as $f=p/q$, where $p$ and $q$ are leading
terms of $P$ and $Q$, respectively. In this case $f(\vxi)$ is a homogeneous
rational function of $\vxi$. To express this fact more precisely we denote by
$\K=\mathscr{M}(\Gamma)$ the field of meromorphic functions on $\Gamma$. Then
$f$ is a rational function of $\xi_1,\ldots, \xi_n$ with coefficients in $\K$,
i.e., $f\in\K(\vxi):=\K(\xi_1,\ldots,\xi_n)$.

It is not difficult to prove that if $F$ is a meromorphic first integral of 
equation~\eqref{eq:ds}, then its leading term $f$ is a first integral of
variational  equations~\eqref{eq:vds}. Moreover, if $F_1, \ldots, F_k$ are
functionally independent meromorphic first integrals of~\eqref{eq:ds}, then, by
the Ziglin Lemma, we can assume that their leading terms $f_1,\ldots, f_k$ are
functionally independent first integrals of~\eqref{eq:vds}. For proofs
and
details see \cite{Ziglin:82::b,Audin:01::c,Churchill:96::b,Morales:99::c}.

Let $\mathscr{G}\subset \mathrm{GL}(n,\C)$ denote the differential Galois group
of~\eqref{eq:vds}. If  $f(g(\vxi))= f(\vxi)$ for every $g\in\mathscr{G}$,
then we say that $f$ is an invariant of $\mathscr{G}$. 
The following lemma says that first integrals of variational equations are invariants of their differential Galois group. It was proved 
in \cite{Morales:99::c}, see also  Lemma
III.1.13 on p.63 in~\cite{Audin:01::c}. 
\begin{lemma}
\label{lem:dgg}
If equation~\eqref{eq:ds} has $k$ functionally independent first integrals which
are meromorphic in a connected neighborhood of  a non-equilibrium solution
$\vvarphi(t)$, then the differential Galois group $\mathscr{G}$ of the
variational equations
along  $\vvarphi(t)$ has $k$ functionally independent  rational invariants. 
\end{lemma}
For a fixed $t\in\C$ (or, for a fixed point $\vvarphi(t)$ on the Riemann surface
$\Gamma$), a  first integral $f$ of the variational equations is an element of
$\C(\vxi)$. The set of all rational invariants of $\mathscr{G}$ is a field
denoted by
$\C(\vxi)^{\mathscr{G}}$.

The  differential Galois group $\mathscr{G}$ of a system of linear equations is
a linear algebraic group, see e.g. \cite{Kaplansky:76::,Put:03::}.
So, in particular,  it is also a Lie group. This fact  allows  to
reformulate  the necessary conditions for the integrability in the language of
Lie algebras. However, passing
from  a Lie group to its Lie algebra, we have to pay something. Namely, we
take into account only the identity component  $\mathscr{G}^0$  of 
$\mathscr{G}$.   In  particular, if  $\mathscr{G}$ is finite, then the Lie
algebra  of $\mathscr{G}$  is  trivial, i.e., it consists of one zero vector.

Let $\mathfrak{g}\subset \mathrm{GL}(n,\C)$ denote the Lie algebra of
$\mathscr{G}$. An
element $Y\in
\mathfrak{g}$ can be considered as a linear vector field: $\vx\mapsto Y(\vx):=
Y\vx$,
for $\vx\in\C^n$. 
We say that $f\in\C(\vx)$ is an integral of $\mathfrak{g}$, iff $Y(f)(\vx)=\rmd
f(\vx)\cdot Y(\vx)=0$, for all $Y\in \mathfrak{g}$. All rational  first
integrals of
the Lie algebra $\mathfrak{g}\subset \mathrm{GL}(n,\C)$ form a field denoted by
$\C(\vx)^{\mathfrak{g}}$.
\begin{proposition}
If $f_1, \ldots, f_k\in\C(\vx)$ are algebraically independent invariants of an
algebraic  group $\mathscr{G}\subset \mathrm{GL}(n,\C)$, then they are
algebraically independent  first integrals
of the Lie algebra $\mathfrak{g}$ of $\mathscr{G}$.
\end{proposition}
\begin{proof}
 Let $X\in \mathfrak{g}$. Then $g(t)=\exp[tX]$ defines for $t\in \R$, $\abs{t}$
small enough, a one parameter subgroup of  $\mathscr{G}$. Thus if $ f\in
\C(\vx)^{\mathscr{G}}$, then
\begin{equation*}
 f(\exp[tX] \vx ) = f(\vx),
\end{equation*}
for any $X\in \mathfrak{g}$. 
Differentiating the above equality with respect to $t$ at $t=0$, we obtain
$X(f)=0$, i.e., $f\in\C(\vx)^{\mathfrak{g}}$.
\end{proof}

It can happen that two systems, one integrable and the other non-integrable,
have the same variational equations along a chosen particular solution. In such
 situation, to show non-integrability of the second system we have to apply
either another method or another particular solution or investigate higher order
variational equations. The last
approach appears very effective. In this paper we use it analyzing several
examples and it is described in Appendix B.
\section{Differential Galois group of variational equations}
\label{sec:part}
We showed in Introduction that the variational equations~\eqref{eq:unve} for a
particular solution~\eqref{eq:part} corresponding to a Darboux point $\vd$ are a
direct
product of second order equations of the form
\begin{equation}
 \label{eq:1ve}
 \ddot \eta = -\lambda \varphi(t)^{k-2} \eta, \qquad \lambda\in\C, 
\end{equation}
where $\varphi(t)$ satisfies~\eqref{eq:hypel}. Let
$\mathscr{G}(k,\lambda)\subset \mathrm{Sp}(2,\C)$ denote the differential
Galois group of the above equation and $\mathscr{G}(k,\lambda)^\circ$ the
identity component of $\mathscr{G}(k,\lambda)$.
Then the differential Galois group $\mathscr{G}$ of variational
equations~\eqref{eq:unve} is
 a direct
product
\begin{equation}
 \label{eq:dirs}
\mathscr{G}= \mathscr{G}(k,\lambda_1)\times \cdots\times
\mathscr{G}(k,\lambda_n)\subset \mathrm{Sp}(2n,\C).
\end{equation} 
Hence, it is clear that in order to know the properties of $\mathscr{G}$, we
should
know as much as possible about $\mathscr{G}(k,\lambda)$. Fortunately, the
properties
of the last group can be described in details.
Following  Yoshida \cite{Yoshida:87::a} it is convenient to define a new
independent variable $z$ in~\eqref{eq:1ve} by
\[
 z :=  \frac{1}{\varepsilon}\varphi(t)^{k},
\]
where $\varphi(t)$ satisfies~\eqref{eq:hypel}.
Then equation~\eqref{eq:1ve} transforms into a  linear
second order equation with rational coefficients
\begin{equation}
\label{eq:hypi}
z(1-z)\eta'' + \left(\frac{k-1}{k} - \frac{3k-2}{2k}z\right)\eta' +
\frac{\lambda_i}{2k} \eta =0.
\end{equation} 
This equation is a special case of 
 the Gauss
hypergeometric equation
\begin{equation}
\label{eq:hyp}
z(1-z)\eta'' +[c-(a+b+1)z]\eta' -ab \eta=0, 
\end{equation} 
with
\begin{equation}
\label{eq:abc}
 a+b=\frac{k-2}{2k},\qquad ab=-\frac{\lambda}{2k},\qquad c=1-\frac{1}{k}.
\end{equation} 
Thus, the differences of exponents at singularities $z=0,1$  and $\infty$  for
equation~\eqref{eq:hyp} are
\begin{equation*}
\rho=1-c=\frac{1}{k}, \qquad \sigma=c-a-b=\frac{1}{2} ,\qquad \tau=a-b=
\frac{1}{2k}\sqrt{(k-2)^2
+8k\lambda},
\end{equation*} 
respectively.
Let $ G(k,\lambda)\subset \mathrm{GL}(2,\C)$ denote a differential Galois
group of~\eqref{eq:hypi}.
The differential Galois group of the hypergeometric equation is well known, see
e.g. \cite{Kimura:69::,Iwasaki:91::}. However, to make this fact usefull, we should know a
relation between
$\mathscr{G}(k,\lambda)$ and $G(k,\lambda)$. It can be shown, see Proposition
4.7 in~\cite{Churchill:96::b}, that the
identity components of these groups are the same.

The well known Kimura theorem~\cite{Kimura:69::} specifies all cases when  all
solutions of the
hypergeometric equation~\eqref{eq:hyp} are Liouvillian.  In these cases, by the
Lie-Kolchin theorem, the identity component of the differential Galois group is
solvable. A direct application of the Kimura theorem to equation~\eqref{eq:hypi}
gives the following.
\begin{lemma}
\label{lem:kim}
The identity component $G(k,\lambda)^\circ$ of the differential Galois group of
equation~\eqref{eq:hypi} with $k>2$ 
is solvable if and only if $(k,\lambda)$  belong to the list given  in
Theorem~\ref{thm:MoRa}.
\end{lemma}
The above lemma is not sufficient for our purpose.  We show the following. 
\begin{lemma}
 \label{lem:ikim}
Assume that $(k,\lambda)$ belongs to the list given in Theorem~\ref{thm:MoRa}
and
$k>2$.
Then, for the differential Galois group $G(k,\lambda)$ of
equation~\eqref{eq:hypi}, the following statements hold.
\begin{enumerate}
 \item $G(k,\lambda)^\circ$ is Abelian. 
\item If $(k,\lambda)$ belongs to the item $1$ of the list, then
 \begin{equation}
 G(k,\lambda)^\circ =\mathscr{T}_1 := \defset{
\begin{bmatrix}
 1 & c\\
0 & 1
\end{bmatrix}
}{c\in\C}.
 \end{equation}
\item If $(k,\lambda)$ belongs to  items $2$--$5$ of the list,  then 
$G(k,\lambda)^\circ =\{E\}$.
\end{enumerate}
\end{lemma}
Here $E$ is the identity matrix.
\begin{proof}
At first we transform equation~\eqref{eq:hypi} into the normal form. To this end
we put
\begin{equation}
\label{eq:ton}
w= \eta \exp\left[\frac{1}{2}\int p\, \rmd z\right], \qquad
p:=\frac{c-(a+b+1)z}{z(1-z)},
\end{equation}
where $a$, $b$ and $c$ are given by~\eqref{eq:abc}.
Then we obtain
\begin{equation}
 \label{eq:nhyp}
w'' = \frac{\rho^2 -1+z(1-\rho^2-\tau^2+\sigma^2) +z^2(\tau^2-1)}{4z^2(z-1)^2}
w.
\end{equation}
For this equation exponents at $0$, $1$ and at infinity are
\begin{equation}
\label{eq:exphypn}
 \left\{  \frac{1}{2}(1-\rho),  \frac{1}{2}(1+\rho) \right\}, \quad  \left\{ 
\frac{1}{2}(1-\sigma),  \frac{1}{2}(1+\sigma) \right\}, \quad  \left\{ -
\frac{1}{2}(1-\tau), - \frac{1}{2}(1+\tau) \right\},
\end{equation} 
respectively. Monodromy and differential Galois groups  of \eqref{eq:nhyp} are
now  subgroups of
$\mathrm{SL}(2,\C)$.  It is important to remark here that the identity
components of the differential Galois group of~\eqref{eq:hypi}
and~\eqref{eq:nhyp} are the same. 
Notice also that the differences of exponents at singular points were
unchanged. 
We denote by $\widehat{G}(k,\lambda)$
 the differential Galois group of equation~\eqref{eq:nhyp}.

By assumption $G(k,\lambda)^\circ$ is solvable, and thus
$\widehat{G}(k,\lambda)^\circ$ is also solvable.
Suppose that $\widehat{{G}}(k,\lambda)^\circ$ is solvable but not Abelian. 
Then, by Theorem 4.12 on p. 31 in \cite{Kaplansky:76::} and Proposition~4.2 
in~\cite{Singer:93::a},
 there is only one
possibility: $\widehat{{G}}(k,\lambda)=
\widehat{{G}}(k,\lambda)^\circ=\mathscr{T}$,
where
$\mathscr{T}$ is the triangular subgroup of $\mathrm{SL}(2,\C)$. So, such case
can appear only if the considered equation is reducible. Using the well known
criterion for the reducibility of Riemann $P$ equation, see
e.g.~\cite{Kimura:69::},
we easily obtain that equation~\eqref{eq:nhyp} is reducible iff $\lambda=
p+kp(p-1)/2$ where $p\in\Z$.

We show that if equation~\eqref{eq:nhyp} is reducible, then
$\widehat{{G}}(k,\lambda)$ is a proper subgroup of $\mathscr{T}$, and in this
way we
 prove the first  statement  of our lemma.

First, let us notice that in the considered reducible case the respective
exponents at singular
points $0$, $1$ and infinity are following
\begin{equation}
\label{eq:expnhr}
 \left\{ \frac{1}{2} -\frac{1}{2k}, \frac{1}{2} + \frac{1}{2k} \right\},
\qquad  \left\{\frac{1}{4}, \frac{3}{4}\right\}, \qquad
\left\{   -\frac{2+k(l+2) }{4k},  \frac{2+k(l-2) }{4k}  \right\},
\end{equation}
where $l$ is an odd integer.
  The difference of exponents at the singular point $z=1$ is $1/2$. Thus,
by Lemma~4.3.6 on p. 90 in~\cite{Iwasaki:91::} 
equation~\eqref{eq:nhyp} has a solution of the form:
\begin{equation*}
 w =z^r(1-z)^s h(z), 
\end{equation*}
where $h(z)$ is a polynomial, and $r$ is an exponent at $z=0$,  and $s$ in an
exponent at $z=1$, As $r$ and $s$ are rational,
there exists $j\in\N$ such that $w^j\in\C(z)$. Now, by Proposition~4.2 
in~\cite{Singer:93::a}, 
 $\widehat{\mathscr{G}}(k,\lambda)$ is either a proper subgroup of
the diagonal group, or a proper subgroup of the triangular group. This finishes
proof of the first statement of the lemma.

To prove the second statement of  we have  to show that
$\widehat{\mathscr{G}}(k,\lambda)$  cannot be a subgroup of the diagonal group.
Assume that it is diagonal. Then, in particular, the monodromy group
of equation~\eqref{eq:nhyp} is diagonal. This last group is generated by  two
elements $M_0$, $M_1\in \mathrm{SL}(2,\C)$--- the monodromy matrices
corresponding to canonical loops encircling singular points $0$ and $1$,
respectively. If the monodromy group of equation~\eqref{eq:nhyp} is diagonal,
then by Lemma~4.3.5 on p. 90 in ~\cite{Iwasaki:91::}, at least one of the
matrices $M_0$, $M_1$ or $M_0M_1$ is a scalar matrix, i.e., $\pm E$. Taking into
account, that the eigenvalues of $M_i$ are $\exp 2\pi \rmi r_{i,1}$ and $\exp
2\pi
\rmi r_{i,1}$, where  $r_{i,1}$ and $r_{i,2}$ are exponents at $z=i$, we easily
conclude that for $k>2$ it is impossible. This finishes the proof of the second
statement.

To show the last statement it is enough to notice that if $(k,\lambda)$ belongs
to item $2$ of the list~\eqref{eq:tab}, then, using Theorem~2.9 (b) on p. 525
in~\cite{Churchill:99::a} one can check that $\widehat{{G}}(k,\lambda)$ is
finite. Moreover, from the proof of the Kimura theorem given
in~\cite{Kimura:69::} 
it follows that if $(k,\lambda)$ belongs to items $3$--$5$ of the
list~\ref{eq:tab}, then ${{G}}(k,\lambda)$, and thus
$\widehat{{G}}(k,\lambda)$, is a finite primitive group.  As the identity
component of a finite group is just the identity element, this finishes the
proof.
\end{proof}
Passing from group $G(k,\lambda)$ to its  Lie algebra from  the lemma proved above we have as an immediate consequence the following 
corollary.   
\begin{corollary}
 \label{cor:ikim}
 Let $\mathfrak{g}$ be the Lie algebra of the differential Galois group
$G(k,\lambda)$ of equation~\eqref{eq:hypi}. Assume that $(k,\lambda)$ belongs to
the list given in Lemma~\ref{lem:kim} and $k>2$. If $(k,\lambda)$ belongs to the
item $1$, then $\dim_{\C}\mathfrak{g}=1$, and otherwise
$\dim_{\C}\mathfrak{g}=0$.
\end{corollary}
\section{Necessary conditions for the Jacobi integrability}
\label{sec:nec}
The main purpose of this section is to give proofs of Theorems~\ref{thm:mu}
and~\ref{thm:MoRa}.
For this purpose we develop appropriate tools investigating certain Lie
algebras.

\subsection{Certain Poisson algebra}
\label{ssec:pa}
An element $Y$ of  Lie algebra
$ \mathrm{sp}(2n,\C)$, considered as a linear vector field, is a Hamiltonian
vector field  given by a global Hamiltonian function $H:\C^{2n}\rightarrow \C$,
which is a homogeneous
polynomial of  $2n$ variables $(x_1,\ldots,x_n,y_1,\ldots,y_n)$ of degree $2$, i.e.  $H\in
\C_2[\vx,\vy]:=\C_2[x_1,\ldots,x_n,y_1,\ldots,y_n]$.   In this way, we identify Lie algebra 
$\mathrm{sp}(2n,\C)$  with a $\C$-linear vector space
$\C_2[\vx,\vy]$ with
the canonical Poisson bracket as the Lie bracket.  Thus, for a Lie algebra
$\mathfrak{g}\subset
\mathrm{sp}(2n,\C)\simeq \C_2[\vx,\vy]$, a rational function $f\in\C[\vx,\vy]$
is a first integral of $\mathfrak{g}$, iff $\{H,f\}=0$, for all $H\in
\mathfrak{g}$.

Now, we consider a more detailed case when $ \mathfrak{g}$ is a Lie subalgebra
of
$\mathrm{sp}(2,\C)$. It is easy to show that Lie algebra $\mathrm{sp}(2,\C)$
does not admit any non-constant first integral.
\begin{proposition}
A rational function  $f\in\C(x,y)$ is a first integral of $\mathrm{sp}(2,\C)$,
iff $f\in\C$.
\end{proposition}
\begin{proof}
Let $f\in \C(x,y)$ be a first integral of $\mathrm{sp}(2,\C)\simeq
\C_2[x,y]$. Thus, $\{f,H\}=0$, for each $H\in\C_2[x,y]$. Let us take $H=x^2$.
Then,
\begin{equation*}
\{f,H\}= -2x\pder{f}{y}=0, 
\end{equation*}
and this shows that $f$ does not depend on $y$, i.e., $f\in\C(x)$.  Taking
$H=y^2$, we show that
$f$ does not depend on $x$. Hence $f\in\C$. 
\end{proof}
The above proposition shows that only  proper subalgebras of $\mathrm{sp}(2,\C)$
can have non-constant first integrals.
\begin{proposition}
\label{prop:dim1}
If $\mathfrak{g}$ is a Lie subalgebra of $\mathrm{sp}(2,\C)$ and $\dim_{\C}
\mathfrak{g}>0$, then the number of algebraically independent rational first
integrals
of $\mathfrak{g}$ is not greater than one.
\end{proposition}
\begin{proof}
As $\dim_{\C} \mathfrak{g}>0$, there exists a non-zero $H\in
\mathrm{sp}(2,\C)\simeq
\C_2[x,y]$. The number of  rational algebraically independent  first integrals
of
a  non-zero linear Hamiltonian vector field $X_H$  in $\C^2$ is  at most one.
\end{proof}
\begin{proposition}
\label{prop:dim2}
If $\mathfrak{g}$ is a Lie subalgebra of $\mathrm{sp}(2,\C)$ and $\dim_{\C}
\mathfrak{g}=2$, then  $\C(x,y)^{\mathfrak{g}}=\C$.
\end{proposition}
\begin{proof}
 All two dimensional Lie algebras are solvable, so  $ \mathfrak{g}$ is solvable.
Thus
a connected Lie group
 $G\subset \mathrm{sp}(2,\C)$ with  Lie algebra  $\mathfrak{g}$ is solvable.  By
the
Lie-Kolchin theorem $G$ is conjugate  to the triangular group
\begin{equation}
\mathscr{T}:=\defset{ \begin{bmatrix}
                      a& b \\
                      0 &a^{-1}
                      \end{bmatrix}}{ a\in\C^\star, \quad b\in\C}.
\end{equation}
The Lie algebra $\mathfrak{t}$ of  $\mathscr{T}$ is isomorphic to
$\mathfrak{g}$, and is generated by two
elements
\begin{equation}
h_1 = \begin{bmatrix}
       1 & 0\\
       0 & -1
      \end{bmatrix}, \qquad
  h_2 = \begin{bmatrix}
       0 & 1\\
       0 & 0
      \end{bmatrix}.
\end{equation}
 Let $H_1$ and $H_2$ be Hamiltonian functions from $\C_2[x,y]$ such that linear
vector fields $X_{H_1}$ and $X_{H_2}$ are represented by matrices $h_1$ and
$h_2$,
respectively. It is easy to check that
\begin{equation}
H_1 = xy, \qquad H_2= \frac{1}{2}y^2.
\end{equation}
We show that  $\C(x,y)^{\mathfrak{t}}=\C$.
Assume that there exists $f\in\C(x,y)^{\mathfrak{t}}\setminus\C$.   Hence
$\{f,H_i\}=0$ for $i=1,2$. But
\begin{equation*}
\{f,H_2\}= y\pder{f}{x} = 0,
\end{equation*}
so, $ f\in\C(y)$. However,  for $ f\in\C(y)$, we have 
\begin{equation*}
\{f,H_1\}= -y\pder{f}{y} = 0,
\end{equation*}
and this implies that $f\in\C$. A contradiction with the assumption that $f$ is
not
a constant shows that $\C(x,y)^{\mathfrak{t}}=\C$.  Moreover, as the Lie
algebras
 $\mathfrak{t}$ and  $\mathfrak{g}$ are isomorphic, we have also
$\C(x,y)^{\mathfrak{g}}=\C$.
\end{proof}
Let $\K\supset\C$ be a field. Then, for the Lie algebra $\mathrm{sp}(2,\C)
\simeq\C_2[x,y]$, we can also consider  first integrals which belong to
$\K(x,y)$.  Here we assume that 
\begin{equation*}
\pder{a}{x}=0\mtext{and}\pder{a}{y}=0\mtext{for all} a\in\K.
\end{equation*}
For our further considerations we need the following lemma.
\begin{lemma}
\label{lem:h}
Let $\mathfrak{g}$ be a one dimensional Lie subalgebra of
$\mathrm{sp}(2,\C)\simeq\C_2[x,y]$. If $f\in\K(x,y)$ is a rational first
integral of  $\mathfrak{g}$, then there
exists a nonzero element $h\in\C[x,y]$ such that
$f\in \K(h)$.
 \end{lemma}
\begin{proof}
 Let us assume that $\mathfrak{g}$ is nilpotent, i.e., $\mathfrak{g}$ is
generated by $H=y^2$. Then we have
\begin{equation*}
 0=\{H, f\}= -2y\pder{f}{x}.
\end{equation*}
Thus, $f$ does not depend on $x$, so for this case we choose $h=y$.

The only other possibility is that  $\mathfrak{g}$ is diagonal, i.e., it
is generated by $H=xy$. First, let us assume that $f$ is a polynomial in
$(x,y)$.  We can write $f$ uniquely as a sum of
homogeneous  components. It is clear that
if $f$  is the first integral  of $H$, then each homogeneous component of $f$ is
also
the first integral of $H$. Thus, let us assume that $f$ is homogeneous of degree
$s$
and let us represent it in the form 
\begin{equation}
 f =\sum_{i=0}^sf_ix^iy^{s-i}, \qquad f_i\in \K \mtext{for} i=1,\ldots, s.
\end{equation}
Then we obtain
\begin{equation*}
 0=\{H,f\}=
\sum_{i=0}^sf_i(s-i)x^iy^{s-i}-\sum_{i=0}^sif_ix^iy^{s-i}=\sum_{i=0}
^sf_i(s-2i)x^iy^{s-i}.
\end{equation*}
Hence, $f_i=0$ for $2i\neq s$, and if for even $s=2r$, $f_r\neq0$, then
$f=f_r(xy)^r$. This implies that every homogeneous, and thus arbitrary
polynomial first integral $f\in \K[x,y]$ of $H$, is an element of
$\K[h]$, where $h=xy$.

Now, assume that $f$ is a rational first integral of $H$. Then we can write
$f=P/Q$ where $P$ and $Q$  are relatively prime polynomials in $\K[x,y]$. Hence
we
have
\begin{equation*}
 0=\{H,P/Q\}=\frac{1}{Q^2}\left( Q\{H,P\}-P\{H,Q\}\right),
\end{equation*}
so $ Q\{H,P\}=P\{H,Q\}$. As $P$ and $Q$ are relatively prime, this implies
that
\begin{equation}
\label{eq:rat}
 \{H,P\} = \gamma P \mtext{and} \{H,Q\}=\gamma Q.
\end{equation}
for a certain $\gamma\in \K[x,y]$.
Comparing degrees of both sides in the above equalities, we deduce that
$\gamma\in \K$. If $\gamma=0$,
then $P$ and $Q$ are polynomial first integrals of $H$, so in this case we have
that $f\in \K(h)$.

We show that the case $\gamma\neq 0$ is impossible. 
Let us assume that $\gamma\neq 0$. It is easy to see that if $P\in \K[x,y]$
satisfies equation
\begin{equation}
\label{eq:dar}
 \{H,P\} = \gamma P,
\end{equation} 
 then its every  homogeneous component
 also satisfies this equation. Thus let us assume
that $P$ is homogeneous of degree $s$.  If we
write 
\begin{equation}
 P=\sum_{i=0}^sP_ix^iy^{s-i}, \qquad P_i\in \K \mtext{for} i=1,\ldots, s,
\end{equation}
then   equation~\eqref{eq:dar} leads to the following equality
\begin{equation}
 \sum_{i=0}^sP_i(s-2i-\gamma)x^iy^{s-i}=0.
\end{equation}
This implies that if coefficient $P_i\neq0$, then $\gamma = s-2i$ and $P= P_i
x^i y^{s-i}$. Thus,  every homogeneous solution of~\eqref{eq:dar}
is a monomial   of the form $P_i x^{i}y^{i+\gamma}$, where $
\gamma$ is a non-zero integer  and $i$ is a non-negative integer such that
$i+\gamma\geq 0$. Thus a non-homogeneous solution of ~\eqref{eq:dar} is a
finite sum
\begin{equation*}
 P = \sum_{i+\gamma>0} p_i x^{i}y^{i+\gamma}.
\end{equation*}
But $Q$ satisfies the same equation~\eqref{eq:dar}, so we also have 
\begin{equation*}
 Q = \sum_{j+\gamma>0} q_j x^{j}y^{j+\gamma}.
\end{equation*}
If $\gamma>0$, then $P$ and $Q$ are not relatively prime, because they have a
common factor $y^{\gamma}$. On the other hand, if $\gamma<0$, then they are 
not  relatively prime, either, because they have a common factor $x$. We have a
contradiction  and this finishes the proof.
\end{proof}

 Let us denote by $\mathfrak{s}$ the Lie algebra which is the
direct sum of $n$ copies of  $\mathrm{sp}(2,\C)$
\begin{equation}
\label{eq:s}
\mathfrak{s}:=
\underbrace{\mathrm{sp}(2,\C)\oplus\cdots\oplus\mathrm{sp}(2,\C)}_{n-\text{times
}},
\end{equation}
 and by 
$\pi_i: \mathfrak{s}\rightarrow \mathrm{sp}(2,\C)$, the projection onto the
$i$-th component of $\mathfrak{s}$, for $i=1,\ldots, n$.
\begin{lemma}
\label{lem:bas}
Let $\mathfrak{g}$ be a Lie subalgebra of $\mathfrak{s}$,   and
$\mathfrak{g}_i=\pi_i(\mathfrak{g})$ for $i=1,\ldots,n$.  Assume that $
\mathfrak{g}$ has $2n -2$ rational algebraically independent first integrals. If
for some $1\leq j
\leq n$, $\dim_{\C}\mathfrak{g}_j>1$, then $\mathfrak{g}_i =\{0\}$, for $i\neq
j$.
\end{lemma}
\begin{proof}
We consider elements of Lie algebra $\mathfrak{s}$ as degree two homogeneous
polynomials
of $2n$ variables $(\vx,\vy)$.   Because $\mathfrak{s}$ is a direct sum of the
form~\eqref{eq:s}, we have
\begin{equation*}
\mathfrak{s}\simeq \bigoplus_{i=1}^n \C_2[x_i,y_i], 
\end{equation*}
and  an element  $H\in \mathfrak{s}$ admits a unique representation
\begin{equation*}
H = \sum_{i=1}^n H_i, \qquad H_i \in \C_2[x_i,y_i], \mtext{for}i=1, \ldots, n. 
\end{equation*}
Without loss of generality, we prove our lemma for  $j=n$. 
Let us assume that $\dim_{\C} \mathfrak{g}_n>1$, and let
$f\in\C(\vx,\vy)\setminus\C$ be a first integral of $ \mathfrak{g}$. Then from
the
proofs of Propositions~\ref{prop:dim1} and \ref{prop:dim2} it follows that $f$
does not depend on $(x_n,y_n)$. Thus, if $ \mathfrak{g}$ has $2n-2$ 
algebraically independent rational first integrals $f_i$, then
\begin{equation}
\label{eq:fi}
f_i\in\C(x_1,\ldots,x_{n-1},y_1,\ldots,y_{n-1})\setminus\C, \qquad i=1,\ldots,
2n-2.
\end{equation} 
Assume that there exists $0<i<n$ such that  $ \mathfrak{g}_i\neq 0$. Then there
exists a non-zero $H\in\C_2[x_i,y_i]$ such that
\begin{equation*}
\{f_j, H\}= \pder{f_j}{x_i}\pder{H}{y_i}-\pder{f_j}{y_i}\pder{H}{x_i}=0,
\mtext{for}j=1,\ldots 2n-2.
\end{equation*}
Hence, 
the  linear Hamiltonian vector field $X_H$ considered as a vector field in
$\C^{2n-2}$ with coordinates $(x_1,\ldots,x_{n-1},y_1,\ldots,y_{n-1})$ has
$2n-2$ first integrals \eqref{eq:fi} and so $X_H=0$. Thus we have a
contradiction
 because for a non-zero  $H\in\C_2[x_i,y_i]$,
$X_H\neq 0$.
\end{proof}
\begin{lemma}
\label{lem:bas2}
Let $\mathfrak{g}$ be an Abelian Lie subalgebra of $\mathfrak{s}$,   and
$\mathfrak{g}_i=\pi_i(\mathfrak{g})$ for $i=1,\ldots,n$.  Assume that $
\mathfrak{g}$ has $2n -2$ rational algebraically independent first integrals.
Then $\dim_{\C}\mathfrak{g}\leq 2$.
\end{lemma}
\begin{proof}
As $\mathfrak{g}$ is Abelian, $\mathfrak{g}_i$ is Abelian, and
$\dim_{\C}\mathfrak{g}_i\leq 1$ for $i=1,\ldots,n$.  Let $\dim_{\C}\mathfrak{g} = s$. Then among
$\mathfrak{g}_i$, exactly $s$ have dimension 1, and the remaining are
zero-dimensional.
We can assume that $\dim_{\C}\mathfrak{g}_i=1$ for $i=1,\ldots, s$,  and
$\dim_{\C}\mathfrak{g}_i=0$  for $i=s+1,\ldots, n$.

Let $f\in\C(\vx)^{\mathfrak{g}}$, then, in particular $f$
is a first integral of $\mathfrak{g}_1$. We can consider $f$ as an element of
$\K(x_1,y_1)$, where $\K=\C(x_2,\ldots,x_n,y_2,\ldots,y_n)$. Then, by
Lemma~\ref{lem:h}, there exists $h_1\in\C(x_1,y_1)$ such that
$f\in\K(h_1)=\C(h_1,x_2,\ldots,x_n,y_2,\ldots,y_n)$.
Repeating successively the above reasoning for $\mathfrak{g}_i$, for
$i=2,\ldots,
s$, we obtain that $f$ is an element of $R:=\C(h_1,\ldots, h_s,
x_{s+1},\ldots,x_n, y_{s+1},\ldots, y_n)$,
where $h_i\in\C(x_i,y_i)\setminus\C$, for $i=1,\ldots,s$. Thus a first integral
of $\mathfrak{g}$ depends on $r= s + 2(n-s)= 2n-s$ variables. This implies that
every set
of $p>r$ elements from $R$ is algebraically dependent. Hence, if
$f_1,\ldots,f_{p}$ are algebraically independent elements of $R$, (i.e.
algebraically independent first integrals of $\mathfrak{g}$), then $p\leq r$.
So, from assumption of our lemma $p=2n-2\leq 2n -s$, and thus $s\leq 2$.
\end{proof}
\subsection{Proofs of Theorems~\ref{thm:mu} and~\ref{thm:MoRa}}
\label{ssec:pr}
 Under assumption of both theorems, the differential Galois group of the
variational
equations is a direct product~\eqref{eq:dirs}. Thus  we also have
\begin{equation}
 \label{eq:dirs0}
\mathscr{G}^\circ= \mathscr{G}(k,\lambda_1)^\circ\times \cdots\times
\mathscr{G}(k,\lambda_n)^\circ\subset \mathrm{Sp}(2n,\C).
\end{equation}
Moreover, the Lie algebra $\mathfrak{g}$ is a direct sum
\begin{equation}
 \mathfrak{g}=\bigoplus_{i=1}^n \mathfrak{g}_i, \qquad
\mathfrak{g}_i\in\mathrm{sp}(2,\C).
\end{equation}
We have already remarked that the identity components of the differential Galois
groups of equations~\eqref{eq:1ve}, \eqref{eq:hypi} and \eqref{eq:nhyp} are the
same. Hence, conclusions of Lemma~\ref{lem:kim} and Lemma~\ref{lem:ikim} are
valid for $\mathscr{G}(k,\lambda)^\circ$.

We prove Theorem~\ref{thm:mu} by contradiction. Let us assume that 
$\mathscr{G}^\circ$
is not Abelian. Then there exists $1\leq i \leq n$
such that $\mathscr{G}(k,\lambda_i)^\circ$ is not Abelian. We have two
possibilities. Either $\mathscr{G}(\lambda_i)^\circ$ is solvable and in this
case it conjugates to the whole triangular group of $\mathrm{Sp}(2,\C)$, or it
is not solvable and then it is  $\mathrm{Sp}(2,\C)$, see Proposition~2.2 in
\cite{Morales:99::c}. In both cases dimension of Lie algebra
$\mathfrak{g}_i$ of $\mathscr{G}(k,\lambda_i)^\circ$  is at least 2. Now, by
Lemma~\ref{lem:bas}, for $r\neq
i$, Lie algebras $\mathfrak{g}_r$ of groups $\mathscr{G}(k,\lambda_r)$
are zero dimensional.

 We know that among
$\lambda_j$ one, let us say $\lambda_s$, $s\in\{1,\ldots, n\}$,  equals
$k-1$, thus it belongs to the item~1 of the table \eqref{eq:tab}. For $\lambda=\lambda_s$ the identity component $G(k,\lambda)^\circ$ of
differential Galois group of hypergeometric equation~\eqref{eq:hypi} is Abelian
but its Lie algebra has dimension one, see Lemma~\ref{lem:ikim}. But 
$G(k,\lambda)^\circ=\mathscr{G}(k,\lambda)^\circ$, so Lie algebra
$\mathfrak{g}_s$ has dimension one. It follows that $s\neq i$. But it is
impossible
because we have shown that  for each  $k\neq i$,  $k\in\{1,\ldots, n\}$, Lie
algebras
$\mathfrak{g}_k$  are zero dimensional. A contradiction finishes the proof of
Theorem~\ref{thm:mu}

Now, we prove Theorem~\ref{thm:MoRa}.
 From Theorem~\eqref{thm:mu} it follows that  $\mathscr{G}^0$  is Abelian and
thus
 $\mathscr{G}(k,\lambda_i)^\circ$ are Abelian for $i=1,\ldots,n$. As
$\mathscr{G}(k,\lambda_i)^\circ= G(k,\lambda_i)^\circ$, we can apply
Lemma~\ref{lem:kim} and we obtain immediately the first statement of
Theorem~\ref{thm:MoRa}. The last statement of this theorem follows from
Lemma~\ref{lem:bas2} and Corollary~\ref{cor:ikim}.

\section{Global analysis}
\label{sec:glo}
The aim of this section is to present a proof of Theorem~\ref{thm:kojot}. The
central role in our considerations plays a theorem proved in
\cite{Guillot:04::}, see also \cite{Guillot:01::}. This theorem is related to
the Kovalevskaya and Painlev\'e 
analysis. To formulate it we recall basic facts.

Let 
\begin{equation}
 \label{eq:g}
  \ve(\vx) :=\sum_{i=1}^nx_i\frac{\partial \phantom{l}}{\partial x_i}, \qquad
\vx =(x_1,\ldots, x_n),
\end{equation}
be the Euler vector field.   A polynomial vector field $\vv$ given
by
\begin{equation}
 \label{eq:vv}
\vv(\vx):= \sum_{i=1}^nv_i(\vx)\frac{\partial \phantom{l}}{\partial x_i}, \qquad
v_i\in\C[\vx], \quad i=1,\ldots, n,
\end{equation} 
is called  homogeneous of degree $p$, iff $L_{\ve}(\vv)=p\vv$, where $L_{\ve}$
denotes
the Lie derivative along the Euler field. Notice that  a  polynomial vector
field $\vv$ is homogeneous of degree $p$ iff $v_i$ are homogeneous and
\begin{equation*}
\deg v_i = p+1\mtext{for}i=1,\ldots,n.
\end{equation*}
With vector field~\eqref{eq:vv} we associate the following polynomial
differential equation
\begin{equation}
\label{eq:vdsg}\Dt \vx =\vv(\vx), \qquad \vx\in\C^n.
\end{equation}
In this context, abusing notation, we consider $\vv$ as an element of
$\C[\vx]^n$. We say that system~\eqref{eq:vdsg} is homogeneous of degree
$p$ iff $\vv$  as a vector field is homogeneous of degree $p$.
\begin{definition}
\label{def:vdar}
We say that a direction $\vd\in\C^n$ is a Darboux point of a homogeneous vector
field $\vv$, or a Darboux point of system~\eqref{eq:vdsg}, iff 
$\ve(\vd)\wedge\vv(\vd)=\vzero$ and $\vv(\vd)\neq \vzero$.
\end{definition}
Thus vector field $\vv$ and the Euler field are parallel at a Darboux point.
This implies that there exists a unique $\rho\in\C^\star$ such that vector field
$\rho\vv-\ve$ has an equilibrium point at $\vd$.
To find all Darboux points of a given vector field $\vv$, it is enough to find
all
non-zero solutions of non-linear equations
\begin{equation}
\label{eq:ndar}
 \vv(\vd)=\vd,
\end{equation} 
and distinguish among them those giving different directions. We always assume
that a Darboux point $\vd$ is normalized in such a way that it
satisfies~\eqref{eq:ndar}. Notice also that a Darboux point
$\vd=(d_1,\ldots,d_n)$ can be considered as a point of the projective space
$\CP^{n-1}$ with homogeneous coordinates $[d_1:\cdots:d_n]$.

The set of all Darboux points of a vector field $\vv$ we denote by $\cD_{\vv}$.
It
can be empty, finite or infinite. One can show, see Proposition~4 in
\cite{Guillot:04::}, that if all Darboux points of a homogeneous vector field
$\vv$ of degree $p$ are isolated, then their number is not greater than
$d(n,p):=((p+1)^n-1)/p$, and  a generic $\vv$ has exactly this number of Darboux
points.

Let $\vd$ be a Darboux point of a homogeneous vector field~\eqref{eq:vv}.   The
Kovalevskaya matrix $K(\vd)$ at  point $\vd$ is defined as
\begin{equation}
 \label{eq:kov}
K(\vd):= \vv'(\vd)-E,
\end{equation} 
and its eigenvalues  are called the Kovalevskaya exponents.
\begin{remark}
 In the above definition we assumed that Darboux point $\vd$
satisfies~\eqref{eq:ndar}. Without this assumption the Kovalevskaya matrix at
Darboux point $\vd$ is defined as the matrix of linearization of vector field
$\rho\vv-\ve$ at equilibrium point $\vd$, i.e.,
\begin{equation*}
 K(\vd):=\rho\vv'(\vd)-E.
\end{equation*}
\end{remark}
We have the following  fact.
\begin{proposition}
\label{prop:p}
Let $\vd\in\C^n$ be a Darboux point of homogeneous vector field $\vv$ of degree
$p$. Then $\vd$ 
is an eigenvector of the Kovalevskaya matrix $K(\vd)$, and the corresponding
eigenvalue is $p$.
\end{proposition}
An easy proof of the above proposition we leave to the reader.

Let $\Lambda_i=\Lambda_i(\vd)$ for $i=1,\ldots, n$ denote the Kovalevskaya
exponents at a Darboux point $\vd\in\cD_{\vv}$. By Proposition~\ref{prop:p}, one
of them is $p$. We always assume that $\Lambda_n=p$, and denote by
$\vLambda=\vLambda(\vd)=(\Lambda_1,\ldots,\Lambda_{n-1})\in\C^{n-1}$.
Let  $\tau_i$ for $0\leq i \leq n-1$, denote
the elementary symmetric polynomial in $(n-1)$ variables  of degree $i$. In
paper~\cite{Guillot:04::} it was shown that the values of $\vLambda(\vd)$ taken
at
different Darboux points are not arbitrary, i.e., there exist universal
relations among all  $\vLambda(\vd)$, $\vd\in\cD_{\vv}$.
\begin{theorem}
\label{thm:Rkojot}
Assume that a homogeneous polynomial vector field $\vv$ of degree $p>0$ has
$d(n,p)$ isolated Darboux points,
 and let $S$ be a symmetric homogeneous polynomial  in
$n-1$ variables of degree 
lower than  $n$. Then, the number   
\begin{equation}
\label{eq:Rkojot}
R:
=\sum_{\vd\in\mathcal{D}_{\vv}}\frac{S(\vLambda(\vd))}{\tau_{n-1}(\vLambda(\vd))
},
\end{equation}
depends only on a choice of $S$, the dimension $n$ and  the degree $p$.
\end{theorem}
In other words, $R=R(S,n,p)$ does not depend on a specific choice of $\vv$,
provided that $\vv$ has the maximal number of Darboux points $d(n,p)$.  For
applications
of the above theorem we have to know values $R(S,n,p)$ for an arbitrary $n$,
$p$, and
for a chosen set of independent symmetric  polynomials $S$ of degree not greater
than $n-1$. The standard choice for $S$  are elementary symmetric polynomials
$\tau_i$, or alternatively $\tau_1^i$, for $0\leq i\leq n-1$.
 To calculate  $R(S,n,p)$ it is enough to find a  homogeneous  system defined
for an arbitrary $n>2$ and  $p\geq 0$ possessing the maximal number of Darboux
points, and such that one can determine the Kovalevskaya exponents easily for
all its
Darboux points.  These requirements are satisfied be a
$n$-dimensional generalization of the Jouanolou system \cite{Maciejewski:00::b}
\begin{equation}
 \label{eq:ju}
 \dot x_i = x_{i+1}^{p+1},  \qquad 1\leq i\leq n, \quad x_{n+1}\equiv x_i.
\end{equation}
For this system, the Kovalevskaya exponents at
all Darboux points are the same and are given by
\begin{equation}
 \Lambda_i = (p+1) \varepsilon^{n-i} -1, \qquad  1\leq i\leq n,
\end{equation}  
where $\varepsilon$ is a primitive $n$-th root of the unity.  
In~\cite{mp:06::f} we have shown that
\begin{equation}
 \label{eq:taur}
 \tau_r(\vLambda)=(-1)^r \sum_{i=0}^{r}
\binom{n-i-1}{r-i}(p+1)^{i},
\end{equation} 
for $ 0\leq r \leq n-1$.  Using this fact one can show that
\begin{equation}
\label{eq:Rt1rp}
 R(\tau_1^r,n,p) = (-1)^{n+r-1}(n+p)^r,
\end{equation}
and
\begin{equation}
 \label{eq:rksymp}
 R(\tau_r,n,p)   = (-1)^{n-i-1}\sum_{i=0}^{r}\binom{n-i-1}{r-i}(p+1)^{i},
\end{equation} 
for $0\leq r\leq n-1$.

Now we return to the Newton equations~\eqref{eq:newton1}. We associate with them
the
following auxiliary system
\begin{equation}
 \label{eq:ps}
  \Dt \vx = \vF(\vx), \qquad \vx\in\C^n.
\end{equation}  
As $\deg F_i=k-1$ for $i=1,\ldots n$,  $\vF$ considered as a vector field has
degree $p=k-2$.  A Darboux point of force $\vF$ is a Darboux point of the
auxiliary system. Thus, if $\vd$ is a Darboux point of~\eqref{eq:ps}, then among
the Kovalevskaya exponents calculated at  $\vd$,  one is  $k-2$.  
We denote the remaining ones  by $\vLambda(\vd)=(\Lambda_1(\vd),\ldots,
\Lambda_{n-1}(\vd))$.
Assuming that the auxiliary system has $d(n,k-2)=D(n,k)$ isolated Darboux points
we can apply Theorem~\ref{thm:Rkojot}. Then, taking into account
relations~\eqref{eq:Rt1rp} and~\eqref{eq:rksymp}, we obtain the thesis of
Theorem~\ref{thm:kojot} and this finishes its proof.

One of the most important consequences of Theorem~\ref{thm:kojot} is the
following.
Let $\cF(n,k)$ denote a set  of forces with  exactly $L=D(n,k)$ Darboux points
$\vd_j$, $j=1,\ldots, L$.  For each $\vF\in\cF(n,k)$ we have $\kappa=(n-1)L$ 
nontrivial Kovalevskaya exponents. We denote their collection by
$\cK(\vF)\in\C^{\kappa} $,
i.e.,
\begin{equation*}
 \cK(\vF):=(\vLambda(\vd_1), \ldots,\vLambda(\vd_L)).
\end{equation*}
If for $\cK(\vF)$ given above,  all  $(k, \lambda_{i,j})$,
$\lambda_{i,j}=\Lambda_i(\vd_j)+1$ for $i=1,\ldots,n-1$ and $j=1,\ldots,L$ 
belong to table~\eqref{eq:tab} in Theorem~\ref{thm:MoRa}, then  we write
$\cK(\vF)\in\cJ(n,k)$. In other words, if for $\vF\in\cF(n,k)$ the corresponding
Newton's equations are integrable in the Jacobi sense, then $\cK(\vF)\in
\cJ(n,k)$, and if $\cK(\vF)\in \cJ(n,k)$, then the Newton equations satisfy the
necessary conditions for the integrability in the Jacobi sense of
Theorem~\ref{thm:MoRa}. We show that sets $\cJ(n,k)$ are finite.
\begin{theorem}
 \label{thm:my}
For $n$, $k\geq 2$ set $\cJ(n,k)$ is finite.
\end{theorem}
\begin{proof}
 Relation \eqref{eq:rksym} in Theorem~\ref{thm:kojot} for $r=n-2$  and $p=k-2$
reads
\begin{equation}
 \label{eq:invL}
 \sum_{j=1}^L\sum_{i=1}^{n-1}\frac{1}{\Lambda_i(\vd_j)}=
-\frac{(k-1)^n-n(k-2)-1}{(k-2)^2}.
\end{equation}
By definition of  $\cJ(n,k)$, quantities $\lambda_{i,j}=\Lambda_i(\vd_j)+1$  for
$i=1,\ldots,n-1$ and $j=1,\ldots,L$  belong to sets listed in
table~\eqref{eq:tab} in Theorem~\ref{thm:MoRa}. Because of this, we can apply
directly Lemma~B.1 from~\cite{Maciejewski:05::b} and this finishes the proof.
See also the proof of Theorem~4 in  \cite{mp:06::f}.
\end{proof}

\section{ Two degrees of freedom}
\label{sec:appgen}
In this section we investigate in detail the case of the Newton equations with
two
degrees of freedom. The main aim is to show that in this case we can obtain 
more general results than those given by Theorem~\ref{thm:kojot}.
Moreover, we investigate also the case of forces which admit infinitely many
Darboux
points, as well as cases when forces do not have Darboux points.

Before we start our analysis of the system with two degrees of freedom,  we make
 two general remark.  If a force
admits Darboux points $\vd_1,\ldots,\vd_l$, $1\leq l\leq n$, then we can  choose
 $l$ coordinate axes along them.  In a such choosen frame  in such   all but $j$-th coordinates of
$\vd_j$  are zero.   

Newton's system of the form~\eqref{eq:newton1} contains as a
subclass natural Hamiltonian systems. Thus it is convenient to have a simple
criterion to distinguish such systems.  In  Appendix A we present two methods
that 
enable to check the equivalence of the analyzed Newton system to a Hamiltonian
one.

In the rest of this section we  assume that $\vF=(F_1, F_2)$, and we put
\begin{equation}
F_j=\sum_{i=0}^{k-1}F_{i}^{j}q_1^{k-1-i}q_2^{i},\mtext{for} j=1,2.
\label{eq:dwa}
\end{equation}
\subsection{Darboux Points}

 Darboux points of a force $\vF$ are the equilibria of the vector field $
\vG(\vq)=\vF(\vq)-\vq$. Thus,   it is convenient to introduce an
auxiliary system
\begin{equation}
  \label{eq:aux}
   \dot q_1 = -q_1 + F_1, \qquad
\dot q_2 = -q_2 + F_2.
\end{equation}
However, we must remember that different equilibria of the above system can
correspond to the same Darboux point. To cope with this problem we introduce
coordinates $x=q_1$ and $ z = q_2/q_1$ and define
\begin{equation}
 \label{eq:fi1}
f_i(z):=F_i(1,z) \mtext{for} i=1,2.
\end{equation} 
Then the auxiliary  system~\eqref{eq:aux} in these coordinates  takes the form
\begin{equation}
  \label{eq:auxz}
  \dot x = - x + x^{k-1}h(z), \qquad \dot z = x^{k-2} g(z) ,
\end{equation}
where 
\begin{equation}
  \label{eq:hg}
 h(z) = f_1(z), \qquad g(z)= f_2(z)-zf_1(z).
\end{equation}
Notice that $\deg g \leq k$. 
As it has been mentioned, a Darboux point is a point in a projective space, and
here it
is $\CPOne$ with $[q_1:q_2]$ as homogeneous coordinates.  Hence, $z$ is a
coordinate on the affine part of $\CPOne$.  An equilibrium of~\eqref{eq:auxz} is
a solution of
\begin{equation}
\label{eq:dr}
x^{k-2}h(z)=1,  \qquad x^{k-2}g(z)=0,  \qquad x\neq 0,
\end{equation}
and thus the Darboux points are roots of polynomial $g(z)$ such that $h(z)$ does
not
vanish on them. This immediately implies that  either the number of Darboux
points is infinite when $g=0$, or it is finite and not greater than $k$. We say
that a Darboux point is multiple iff its $z$ coordinate is a multiple root of
$g(z)$.
\begin{remark}
\label{rem:nod}
As it has already been mentioned, not all roots of $g(z)$  are  Darboux points.
If $g(z_\star)=h(z_\star)=0$, then
$z_\star$ is not a Darboux point. Equations $g(z_\star)=h(z_\star)=0$ imply that
 $f_1(z_\star)=f_2(z_\star)=0$, i.e., $F_1$  and $F_2$ have a common factor
$P=q_2-z_\star q_1$.
\end{remark}

If $[0:1]\in\CPOne$ is a Darboux point, then we say that it is located at the infinity. To
investigate its neighborhood  we use coordinates
$y=q_2$ and $\zeta=q_1/q_2$.  We define
\begin{equation}
\label{eq:tfi}
 \tilde f_i(\zeta):=F_i(\zeta,1) ,\mtext{for} i=1,2,
\end{equation} 
and 
\begin{equation}
 \label{eq:tgh}
\tilde g(\zeta):=\tilde f_1(\zeta)-\zeta \tilde f_2(\zeta), \qquad \tilde
h(\zeta):=\tilde f_2(\zeta).
\end{equation}
Then a Darboux point is a root of polynomial $\tilde g$, such that it is not a
root of $\tilde h$.  If $\tilde g(0)=\tilde f_1(0)=0$ and simultaneously $\tilde
h(0)=\tilde f_2(0)\neq 0$, then the
infinity is a Darboux point. Notice that
\begin{equation}
\tilde f_i(\zeta) =\zeta^{k-1} f_i\left(\frac{1}{\zeta}\right) \mtext{for}i=1,2.
\end{equation}

In many cases, it is convenient to assume that all Darboux points are located in
the affine part of $\CPOne$, i.e., that there is no Darboux point at the infinity.
If $\deg g =k$, then surely the infinity is not a Darboux point. Hence, let us
analyze whether we can assume that for a given force the corresponding
polynomial $g$ is of degree $k$.  From the definition of polynomial $g$,
see~\eqref{eq:hg}, it follows that $\deg g = k$ if and only if $\deg f_1=k-1$.
We show the following.
\begin{proposition}
 \label{prop:1>2}
Let $\vF=(F_1,F_2)$, $F_1,F_2\in\C[q_1,q_2]$ be a homogeneous force and $\deg
F_i=k-1$ for $i=1,2$. Assume that for each  $ A\in \mathrm{GL}(2,C)$  for the
equivalent force $\vF_{A}=( F_1^A, F_2^A)$  we have $\deg_{q_2}  F_1^A< k-1$.
Then
\begin{equation}
\label{eq:G}
 G := q_1F_2 -q_2F_1 =0.
\end{equation} 
\end{proposition}
\begin{proof}
One can directly check that if $\vF$ satisfies condition~\eqref{eq:G}, then
every force equivalent to $\vF$ also satisfies this condition.

First let us notice that condition $\deg_{q_2}  F_1< k-1$ is equivalent to $\deg
f_1<k-1$, and this is equivalent to $\tilde f_1(0) =0$.  In fact, if
\begin{equation*}
 f_1(z)=\sum_{i=0}^lf_i^1z^i,\qquad f_l^1\neq 0 \mtext{and}l<k-1,
\end{equation*}
then
\begin{equation*}
 \tilde f_1(\zeta) =
\zeta^{k-1}f_1\left(\frac{1}{\zeta}\right)=\sum_{i=0}^lf_i^1
\zeta^{k-1 -i},
\end{equation*}
and hence $\tilde f_1(0)=0$. 

Let
\begin{equation*}
 A = \begin{bmatrix}
        \alpha & \beta\\
        \gamma & \delta
       \end{bmatrix}, \qquad \Delta = \alpha\delta +\beta\gamma\neq 0.
\end{equation*}
Then
\begin{equation}
 F_1^A(q_1,q_2) =\frac{1}{\Delta}\left[\delta F_1(\alpha q_1+\beta q_2,\gamma
q_1 +\delta q_2) -\beta F_2(\alpha q_1+\beta q_2,\gamma q_1 +\delta q_2)\right],
\end{equation}
and hence
\begin{equation*}
 \tilde{f}_1^A(\zeta)= \frac{1}{\Delta}(\gamma \zeta+\delta)^{k-1} \left[
\delta \tilde f_1 (\tau(\zeta))  -\beta \tilde f_2 (\tau(\zeta)) \right],
\end{equation*}
where
\begin{equation*}
 \tau(\zeta) = \frac{\alpha\zeta+\beta}{\gamma\zeta+\delta}.
\end{equation*}
Assume that $\delta\neq 0$ and denote $x=\beta/\delta$. Then we obtain
\begin{equation*}
 \tilde{f}_1^A(0)=\frac{\delta^k}{\Delta}\left[
 \tilde f_1 (x)  -x \tilde f_2 (x) \right] = \frac{\delta^k}{\Delta} \tilde
g(x).
\end{equation*}
By assumption  $\tilde{f}_1^A(0)=0$  for an arbitrary $A$, so
\begin{equation}
  \tilde f_1 (x) - x \tilde f_2 (x)=0 ,
\end{equation}  
for an arbitrary $x\in \C$. Putting $x=q_1/q_2$  in  the above equality an using
the
definition of $\tilde{f}_1$ and  $\tilde{f}_2$, see~\eqref{eq:tfi},  we obtain
condition~\eqref{eq:G}.
\end{proof}
\begin{remark}
 Condition~\eqref{eq:G} means that force $\vF=(F_1,F_2)$ admits infinitely many
Darboux points.
\end{remark}

\subsection{Generic case }
Now we determine the Kovalevskaya exponents at a given Darboux point.
The Jacobian matrix of the auxiliary system~\eqref{eq:auxz} has the form
\begin{equation}
\label{eq:hess} 
  J = \begin{bmatrix}
 -1+(k-1)x^{k-2}h(z) & x^{k-1}h'(z) \\
  (k-2)x^{k-3}g(z) & x^{k-2}g'(z)
\end{bmatrix}.
\end{equation}
For a Darboux point $(x,z)$  satisfying~\eqref{eq:dr} this matrix becomes
\[
  J = \begin{bmatrix}
 k-2 &  xh'(z)/h(z) \\
  0 &  g'(z)/h(z)
\end{bmatrix}.
\]
Hence the non-trivial Kovalevskaya exponent at a Darboux point  with coordinate
$z$
is given by
\begin{equation}
  \label{eq:lz}
   \Lambda(z) =
\frac{g'(z)}{h(z)}=\frac{f_2'(z)-f_1(z)-zf_1'(z)}{f_1(z)}.
\end{equation}

On the projective line $\CPOne$, where Darboux points are located,  we
define a meromorphic differential form $\omega$ which in coordinate $z$
is given by
\begin{equation}
\label{eq:dt} 
\omega = \frac{h(z)}{g(z)}\,\rmd z, 
\end{equation}
and in coordinate $\zeta = 1/z$ in the neighborhood of infinity by 
\begin{equation}
\label{eq:dtzeta} 
\omega_{\infty} = -\frac{h(1/\zeta)}{g(1/\zeta)}\,\frac{\rmd \zeta}{\zeta^2}. 
\end{equation}
Notice that if $z_\star$ is a Darboux point, then it is a pole of 
$\omega$, and the residue of $\omega$ at this point is
 \begin{equation}
  \res(\omega,z_\star) =\frac{h(z_\star)}{g'(z_\star)}
=\frac{1}{\Lambda(z_\star)},
 \end{equation} 
provided that $g$ and $h$ are relatively prime and $z_\star$ is a simple root of
$g$.

For a given force $\vF=(F_1,F_2)$ with finitely many Darboux points, we can
always assume that all of them  are located in the affine part of $\CPOne$ i.e.
for all Darboux points
$q_1\neq 0$. This can be achieved by a proper choice of coordinates. Under this
assumption we have $F_1(0,q_2)\neq 0$, and hence $F^1_{k-1}\neq 0$,
see~\eqref{eq:dwa}.  We keep this assumption in our further considerations.

To formulate a generalization of Theorem~\ref{thm:kojot} we
have to distinguish a class of forces for which this is possible. 
Let $\cA_l$ denote a set of forces  $\vF=(F_1,F_2)$  satisfying the following
conditions
\begin{itemize}
 \item [A.1] $\vF$ admits $1\leq l\leq k$  simple Darboux points;
\item[A.2] if $F_1(\vq)$ and $G(\vq):=q_2F_1(\vq)-q_1F_2(\vq)$ have a common
linear factor $P=\alpha q_1 +\beta q_2$, $\abs{\alpha}+\abs{\beta}\neq 0$, then
the multiplicity of $P$  in $G$ is not greater than the multiplicity of $P$ in
$F_1$.
\end{itemize}

\begin{theorem}
\label{thm:rel}
Assume that $\vF\in\cA_l$, and let $\Lambda_i$  is the non-trivial Kovalevskaya
exponent  of $\vF$ at $i$-th Darboux point for $i=1,\ldots,l$. Then
\begin{equation}
\label{eq:rel}
\sum_{i=1}^l \frac{1}{\Lambda_i} = -1.
\end{equation}
\end{theorem}
\begin{proof}
We can assume that $\deg g(z)=k$ and $\deg h(z)=k-1$. Moreover, from the
definitions of
$h$ and $g$, see~\eqref{eq:hg}, it follows that if $\alpha z^{k-1}$ is the
highest order term of $h$, then $-\alpha z^{k}$ is the highest order term of
$g$.

Although for $l=k$ the statement of theorem coincides with
Theorem~\ref{thm:kojot}, it is instructive to give first an independent proof
for this case. Thus, let $k=l$. Then $g$ and $h$ are relatively prime and all
$k$ roots $z_i$ of $g$ are simple.  We have
\[
\res(\omega,z_i)=\frac{h(z_i)}{g'(z_i)}=\frac{1}{\Lambda_i}\mtext{for}i=1,\ldots
,k.
\]
To calculate the residue of $\omega$  at infinity, i.e.,
$\res(\omega_\infty,0)$,  we notice that for polynomials $\tilde h(\zeta) =
\zeta^{k-1}h(1/\zeta)$ and  $\tilde g(\zeta)=\zeta^{k} g(1/\zeta)$ we have $\tilde
h(0) =\alpha$ and   $\tilde g(0)=-\alpha$. Thus
\begin{equation*}
 -\frac{h(1/\zeta)}{\zeta^2 g(1/\zeta)}=  -\frac{\tilde h(\zeta)}{\zeta \tilde
g(\zeta)}=\frac{1}{\zeta}+\cdots,
\end{equation*}
where dots denote higher order terms. This shows that $\res(\omega_\infty,0)=1$
and the application of the global residue theorem, see e.g. \cite{Griffiths:78::},
 finishes the proof for $l=k$.

For $1\leq l< k$ the assumption that $\vF\in\cA_l$ implies that $g(z)$ has $l$
simple roots $z_1,\ldots,z_l$, such that $h(z_i)\neq 0$ for $i=1,\ldots,l$.
Moreover, if $s\not\in\{z_1,\ldots,z_l\}$ is a root of $g(z)$ with the
multiplicity $j$, then $s$ is also a root of $h(z)$ with the multiplicity greater
or
equal to $j$. All these conditions imply that we can write
\[
g(z)=-\alpha p(z)\tilde g(z),\qquad h(z)=\alpha p(z)\tilde h(z),
\]
where $p(z)$, $\tilde g(z)$ and $ \tilde h(z)$ are monic polynomials; $p$ is of
degree $k-l$, $\alpha\in\C^{\ast}$
and
\[
\tilde g(z)=\prod_{i=1}^{l} (z-z_i).
\]
Hence
\begin{equation*}
 \omega = -\frac{ \tilde h(z)}{\tilde g(z)}\rmd z,
\end{equation*}
and then
\begin{equation}
\label{eq:jresi}
\frac{1}{\Lambda_i}=\frac{h(z_i)}{g'(z_i)}= -
\frac{p(z_i)\tilde h(z_i)}{p'(z_i) \tilde g(z_i) +
 p(z_i) \tilde g'(z_i) }=
-\frac{\tilde h(z_i)}{\tilde g'(z_i)}=  \res(\omega,z_i),
\end{equation}  
for $i=1,\ldots, l$. Moreover, it is easy to see that  $\res(\omega,\infty)= 1$.
 This finishes the proof.
\end{proof}
\begin{remark}
\label{rem:genrel}
If $\vF$ satisfies assumption A.1 and does  not satisfy A.2 we still can do
something. In such situation we have 
\begin{equation}
 g(z)=-\alpha p(z)\tilde g(z), \qquad p(z)=\prod_{i=1}^s(z-s_i)^{n_i}, \quad
\tilde g(z)=\prod_{i=1}^{l} (z-z_i),
\end{equation} 
and 
\begin{equation}
 h(z)=\alpha q(z)\tilde h(z), \qquad q(z)=\prod_{i=1}^s(z-s_i)^{m_i},  
\end{equation} 
where
\begin{equation*}
 l +\sum_{i=1}^s n_i = k, \qquad  \deg \tilde h +  \sum_{i=1}^s m_i =k-1, 
\end{equation*}
$s_i\not\in\{z_1,\ldots, z_l\}$ and $\tilde h(s_i)\neq 0$ for $i=1,\ldots, s$.
Now
\begin{equation}
 \omega = -\dfrac{\tilde h(z)}{\tilde g(z)\prod_{i=1}^s(z-s_i)^{n_i-m_i} }\rmd
z,
\end{equation} 
and the global residue theorem gives
\begin{equation}
\label{eq:genrel}
 \sum_{i=1}^l\frac{1}{\Lambda_i} + \sum_{j=1}^s\res(\omega, s_j)= -1.
\end{equation} 
We notice that if $n_i>m_i$, then, generally, residue  $\res(\omega,
s_i)\neq
0$ and it depends on coefficients of polynomials $F_1$ and $F_2$.  
\end{remark}
\subsection{Non-maximal number of Darboux points and multiple Darboux points}
If a force $\vF=(F_1, F_2)$  does not have  the maximal number of Darboux
points, then polynomials $F_1$ and $F_2$ have a common factor 
$P\in\C[\vq]\setminus\C$, i.e., $F_i=P\widetilde F_i$, where $\widetilde
F_i\in\C[\vq]$ for $i=1,2$.  In such cases it is possible to find a particular
solution which does not correspond to a Darboux point. For example, if $P=q_2$,
then $q_1(t) =\gamma t$, $q_2(t)=0$ is a particular solution. Amazingly, we can use
such  simple solutions and methods of differential Galois theory  for proving
the
non-integrability. First we show a very simple fact.
\begin{proposition}
 \label{prop:rot}
Let $\vF(\vq)=(\alpha_1q_1+\alpha_2q_2)(\widehat F_1(\vq),\widehat F_2(\vq))$
where $\abs{\alpha_1}+\abs{\alpha_2}\neq 0$, and $\widehat F_1$, $\widehat
F_2\in\C[\vq]$. Then there exists $A\in\mathrm{GL}(2,\C)$, such that
$A\vF(A^{-1}\vq)=q_2( \widetilde F_1(\vq),\widetilde F_2(\vq))$, where
$\widetilde F_i\in\C[\vq]$ for $i=1,2$.
\end{proposition}
\begin{proof}
Matrix $A$ given by
\begin{equation*}
 A=\begin{bmatrix}
     \alpha_2^\star & -\alpha_1^\star \\
      \alpha_1 & \alpha_2
     \end{bmatrix}, \qquad \det A = \abs{\alpha_2}^2 +\abs{\alpha_2}^2,
\end{equation*}
is non-singular and satisfies the desired condition. Here asterisk denotes 
the complex conjugation.
\end{proof}
Hence,  if a force $\vF=(F_1, F_2)$ where $\deg F_i=k-1$, has less than $k$
Darboux points, then we can always assume that $F_1$ and $F_2$ have a common
factor $q_2$. We show that under a certain genericity condition the Newton
equations with such force are not integrable in the Jacobi sense.
\begin{theorem}
\label{thm:less}
Assume that $\widetilde{F}_1$, $\widetilde{F}_2\in\C[\vq]$ are homogeneous
polynomials of degree $k-2$.
 If $\deg_{q_1}\tilde{F}_2=k-2$, then system
\begin{equation}
\dot q_1=p_1, \quad
 \dot p_1=-q_2\widetilde{F}_1(q_1,q_2),\qquad  \dot q_2=p_2 ,\quad \dot
p_2=-q_2\widetilde{F}_2(q_1,q_2),
\label{eq:nonmax}
\end{equation}
does not admit two functionally independent polynomial  first integrals.
\end{theorem}
\begin{proof}
System \eqref{eq:nonmax} possesses the following particular solution
\[
 q_1(t)=\gamma t+\beta,\qquad p_1(t)=\gamma,\qquad q_2(t)=p_2(t)=0.
\]
Variational equations along this solution take the form
\begin{equation}
\label{eq:varPQ}
 \begin{pmatrix}
  \dot Q_1\\\dot P_1\\\dot Q_2\\\dot P_2
 \end{pmatrix}=
\begin{pmatrix}
 0&1&0&0\\
 -\widetilde{F}_1(q_1(t)&0&0&0\\
0&0&0&1\\
0&0&-\widetilde{F}_2(q_1(t),0)&0
\end{pmatrix}
 \begin{pmatrix}
   Q_1\\ P_1\\ Q_2\\ P_2
 \end{pmatrix}.
\end{equation}
We have
\[
 \tilde{F}_1(q_1,0)=\alpha_1q_1^{k-2}\mtext{and}\tilde{F}_2(q_1(t),0)=\alpha_2
q_1^{k-2},
\]
where $\alpha_2\neq0$, as, by assumption $\deg_{q_1}\tilde{F}_2(q_1,0)=k-2$. 
We choose constants of integration $\beta=0$ and $\gamma$ in such a way that
 $\alpha_2\gamma^{k-2}=-1$.
Then
 the variational equations  take the form
\begin{equation}
\label{eq:Q2}
\dot Q_1 = P_1,\quad  \dot P_1 = \alpha t^m Q_2, \qquad  \dot Q_2=P_2, \quad 
\dot P_2=t^mQ_2,\qquad \alpha\in\C, \quad m=k-2>0.
\end{equation}
First we show that the differential Galois group $G$ of the above variational
equations is a direct product $\{E\}\times \mathrm{SL}(2,\C)$, 
where  $E$ is the
$2\times 2$ identity matrix. To this end, we notice that $Q_2$ satisfies
 the
following equation
\begin{equation}
\label{eq:2Q2}
 \ddot Q_2 = t^m Q_2.
\end{equation} 
Let $K\supset \C(t)$ be the Picard-Vessiot extention for this equation, i.e.,
$K$ is the smallest differential field containing two linearly independent
solutions of \eqref{eq:2Q2}. Then all solutions of~\eqref{eq:Q2}  belong to
$K^4$. It is clear for $\alpha =0$. If $\alpha\neq 0$, then $\ddot Q_1 -\alpha
\ddot Q_2 =0$, and so  $Q_1(t) = \alpha Q_2(t) +a t +b$,  for some $a,b\in\C$.
So, as we claimed $ (Q_1(t), \dot Q_1(t), Q_2(t), \dot Q_2(t))\in K^4$.

Now we can determine the differential Galois group of variational
equations~\eqref{eq:Q2}.  For this purpose we have to know the
differential Galois $\widetilde G$ group of equation~ \eqref{eq:2Q2}. In~\cite{Maciejewski:05::b} it was shown that  $\widetilde G=\mathrm{SL}(2,\C)$ for
an arbitrary  $m>0$.  Let $w_1$ and $w_2$ denote two linearly independent
solutions of~ \eqref{eq:2Q2}. Then the fundamental matrix of
\eqref{eq:Q2} has the following form
\begin{equation}
 \label{eq:X}
X = \begin{bmatrix}
 1  & t & \alpha w_1  & \alpha w_2 \\
 0  & 1 & \alpha \dot w_1 & \alpha \dot w_2 \\
0  & 0  & w_1 & w_2 \\
 0 & 0  & \dot w_1 & \dot w_2
\end{bmatrix}.
\end{equation}
Now, if for $ g\in \widetilde G$ we have
\begin{equation}
g(w_i) = \sigma_{1,i}w_1 + \sigma_{2,i}w_2 \mtext{for} i=1,2,
 \end{equation}
then $\Sigma=[\sigma_{i,j}]\in \mathrm{SL}(2,\C)$, and
\begin{equation}
 g(X) = XS, \qquad S = \begin{bmatrix}
 E & 0\\
0 & \Sigma
\end{bmatrix}, 
\end{equation}
where $E$ is the $2\times 2$ identity matrix.  Hence the differential Galois
group $
G$ of system~\eqref{eq:Q2} is a direct product $\{E\}\times \mathrm{SL}(2,\C)$.

Let us notice that we cannot finish our proof at this point invoking
Lemma~\ref{lem:dgg}. In fact, the differential Galois group $G$ has two
independent invariants $Q_1$ and $P_1$.
Thus we proceed in the following way. 
From the proof of Lemma~\ref{lem:bas} it follows that if
$R\in\C(Q_1,P_1,Q_2,P_2)$
is an invariant of $G$, then $R\in\C(Q_1,P_1)$. This implies that if
$I\in\C(t)(Q_1,P_1,Q_2,P_2)$ is a first integral of variational
equations~\eqref{eq:Q2}, then $I\in\C(t)(Q_1,P_1)$.

Let us assume that $P\in \C[q_1,p_1,q_2, p_2]\setminus\C$ is a first integral of
\eqref{eq:nonmax}. We can write it in the following form
\begin{equation}
\label{eq:P}
 P = \sum_{0\leq i+j\leq m}P_{i,j}(q_1,p_1)q_2^ip_2^j.
\end{equation}
The leading term $p\in C[t](Q_1,P_1,Q_2,P_2)$ of $P$ is the lowest degree term
of
\begin{equation}
 P(\gamma t+Q_1,\gamma +P_1, Q_2,P_2) = \sum_{0\leq i+j\leq m}P_{i,j}(\gamma t+Q_1,\gamma
+P_1)Q_2^iP_2^j,
\end{equation}
and it is a first integral of variational equations~\eqref{eq:Q2}. We have
already shown that the integral $p$ must belong to $ C[t](Q_1,P_1)$. Hence $p$
is the lowest order term of  $P_{0,0}(\gamma t+Q_1,\gamma +P_1)$.

We show that $P_{0,0}\in \C[p_1]$.  Inserting expansion~\eqref{eq:P} into
equation
\begin{equation}
 p_1\pder{P}{q_1} + p_2\pder{P}{q_2}-q_2\widetilde F_1(q_1,q_2)\pder{P}{p_1}
-q_2\widetilde F_2(q_1,q_2)\pder{P}{p_2} = 0,
\end{equation}
obtained from the condition $\dot P=0$, 
and taking terms of degree zero with respect to $(q_2,p_2)$,  we obtain
\begin{equation*}
 p_1\pder{P_{0,0}}{q_1}=0, 
\end{equation*}
and hence $P_{0,0}\in\C[p_1]$, as we claimed.

Now, assume that system~\eqref{eq:nonmax} admits two independent polynomial
first integrals $P$ and $Q$. We can assume that their leading terms $p,q \in
C[t](Q_1,P_1,Q_2,P_2)$ are independent.  But we know that  $p,q \in
C[t](Q_1,P_1)$,
and thus, $p$ and $q$ are the lowest degree terms of
$P_{0,0}(\gamma t+Q_1,\gamma +P_1)$ and $Q_{0,0}(\gamma t+Q_1,\gamma +P_1)$. However, we have 
already shown
that $P_{0,0}$ and $Q_{0,0}$ depend only on $p_1$, so $p$ and $q$ depend only on
$P_1$, and thus they are functionally dependent. A contradiction finishes the
proof.
\end{proof}
\begin{remark}
 \label{rem:a}
If in the above lemma we assume that  $\deg_{q_1}F_1 =\deg_{q_1}F_2 =k-2$, then
$\alpha$ in variational equations~\eqref{eq:Q2} is not zero, and we can modify
the final step of the proof. Let $P\in\C(t)(Q_1,P_1)$ be a first integral of
variational equations~\eqref{eq:Q2}. Then the condition $\dot P=0$ yields
\begin{equation*}
 \pder{P}{t}+P_1\pder{P}{Q_1} +\alpha t^m Q_2 \pder{P}{P_1}=0,
\end{equation*}
and thus $ \pder{P}{P_1}=0$.  This implies that the variational equations admit
at most one rational first integral, and thus the considered
system~\eqref{eq:nonmax} is not integrable in the Jacobi sense with two rational
first integrals.
\end{remark}
\begin{remark}
\label{rem:b}
Let us assume that  $\vF(\vq)=(\alpha_1q_1+\alpha_2q_2)(\widehat
F_1(\vq),\widehat F_2(\vq))$.  Without loss of generality we can  assume
that $\alpha_2\neq0$.  If $\alpha_2=0$, we just renumber the coordinates. By
Proposition~\ref{prop:rot} there exists a non-singular matrix $A$ such that
\begin{equation*}
\vF_{A}(\vq):= A\vF(A^{-1}\vq)=q_2( \widetilde F_1(\vq),\widetilde
F_2(\vq)).
\end{equation*}
Assumption $\deg_{q_1}\widetilde F_2 =k-2$ in Theorem~\ref{thm:less} means that
 $\widetilde F_2(q_1,0)\neq 0$.  We have
\begin{equation}
\widetilde F_2(q_1,q_2)=\alpha_1 \widehat F_1(\alpha_2q_1+\alpha_1^\star q_2,
-\alpha_1q_1 +\alpha_2^\star q_2)+\alpha_2 \widehat
F_2(\alpha_2q_1+\alpha_1^\star q_2, -\alpha_1q_1 +\alpha_2^\star q_2),
\end{equation} 
and thus
\begin{equation}
\label{eq:tf2}
 \widetilde F_2(q_1,0)=\alpha_2^{k-1}q_1^{k-2}\left [  \alpha_1 \widehat
F_1(\alpha_2, -\alpha_1 )+ \alpha_2\widehat F_2(\alpha_2, -\alpha_1 ) \right].
\end{equation} 
The above shows that if the polynomial 
\begin{equation*}
 \widehat G(\vq):= q_1\widehat F_2(\vq) -q_2 \widehat F_1(\vq),
\end{equation*}
is divisible by $(\alpha_1q_1+\alpha_2q_2)$, then  $\widetilde F_2(q_1,0)= 0$.
On the other hand, if $\widetilde F_2(q_1,0)= 0$, then  from \eqref{eq:tf2} it
follows that
\begin{equation}
 \hat g(\alpha):= \widehat F_2(1,\alpha) -\alpha  \widehat F_1(1,\alpha)= \hat
f_2(\alpha)-\alpha \hat f_1(\alpha) =0 , \qquad \alpha=
-\frac{\alpha_1}{\alpha_2}.
\end{equation} 
Thus $\alpha$ is a root of polynomial $\hat g\in\C[z]$, and this is equivalent
to the fact that  $(\alpha_1q_1+\alpha_2q_2)$ is a factor of $ \widehat G(\vq)$.
\end{remark}
It can happen that the considered force does not have Darboux points.  Such
forces are characterized in the following theorem.
\begin{theorem}
Assume that a force $\vF=(F_1,F_2)$, $\deg F_i=k-1$, does not have a Darboux
point.  Then
\begin{equation}
\label{eq:bez}
F_1=\prod_{i=1}^{k-1}(q_2-z_iq_1), \qquad
F_2=\frac{1}{q_1}\left[q_2 F_1
-\prod_{i=1}^p(q_2-z_iq_1)^{n_i}\right],
\end{equation}
where  $1\leq p \leq k-1$,  $z_i\in\C$, and $z_1, \ldots, z_p$ are pairwise
different; $n_i\in \N$ and $n_1+\cdots+n_p=k$. Moreover, if  there exist $1\leq
j \leq p$ such that $n_j=1$, then the Newton equations with force $\vF$ do not
admit two functionally independent polynomial first integrals.
\label{thm:g1}
\end{theorem}
Let us note that in spite of the fraction in the expression on $F_2$ the force
$\vF$ 
is polynomial, see the proof below.
\begin{proof}
Without loss of generality we can assume that $\deg_{q_2}F_1=k-1$. Then for
polynomials $h$, $g\in\C[z]$ defined by~\eqref{eq:hg}  we have $\deg g=k$ and
$\deg
h=k-1$. Assumption that $\vF$ have no Darboux point implies that all roots
of $g$ are also roots of $h$.  So $g$ has at least one multiple root and we can
write it in the following form
\[
g(z)=\alpha\prod_{i=1}^p(z-z_i)^{n_i},\qquad n_i\in\N,\quad \sum_{i=1}^pn_i=k,
\quad \alpha \in \C^\star,
\]
where $1\leq p\leq k-1$ and $z_1,\ldots,z_p$ are pairwise different 
roots of $h$. Let $z_{p+1}, \ldots, z_{k-1}$ be the remaining (not necessarily
different) roots of $h$. Then we can write
\[
h(z)=\beta\prod_{i=1}^{k-1}(z-z_i).
\]
 Thus $f_1(z)=h(z)$,  and
 \[
 f_2(z)=g(z)+zh(z)=\alpha \prod_{i=1}^p(z-z_i)^{n_i} 
+\beta z\prod_{i=1}^{k-1}(z-z_i).
 \] 
 But $\deg f_2(z)<k$,  thus $\alpha=-\beta$ and we can put $\beta=1$. 
 Now, as
$F_i(q_1,q_2)=q_1^{k-1}f_i(q_2/q_1)$ for $i=1,2$, we obtain~\eqref{eq:bez} 
and
this finishes the first statement of our theorem.

To proof the second statement let us assume that $n_1=1$. Then we can write
\begin{equation}
 F_i = (q_2 -z_1q_1) \widehat F_i \mtext{for}i=1,2,
\end{equation}
where
\begin{equation}
 \widehat F_1=\prod_{i=2}^{k-1}(q_2-z_iq_1) \qquad \widehat F_2
=\frac{1}{q_1}\left[q_2 \widehat
F_1 -\prod_{i=2}^p(q_2-z_iq_1)^{n_i}\right],
\end{equation}
Let us calculate polynomial $\hat g(z)$ as in Remark~\ref{rem:b}. We have
\begin{equation}
\hat g(z)=\widehat G(1,z) =\widehat F_2(1,z)-z\widehat
F_1(1,z)=-\prod_{i=2}^p(z-z_i)^{n_i},
\end{equation}
so $g(z_1)\neq 0$. Thus, as we explained in  Remark~\ref{rem:b},  one can apply
Theorem~\ref{thm:less} to the considered case and this finishes the proof.
\end{proof}
Let us notice that if the assumptions of the above theorem are satisfied, then
the
considered Newton's system still can possess a single  first integral.  Let
\[
F_1(q_1,q_2)=\prod_{i=1}^{k-1}(q_2-z_iq_1) \mtext{and}
F_2(q_1,q_2)=z_jF_1(q_1,q_2),
\]
where  $1\leq j\leq k-1$ and $z_1,\ldots, z_{k-1}$ are pairwise different. This
force has the form~\eqref{eq:bez} with $p=k-1$, $n_i=1$ for $i\neq j$ and
$n_j=2$. However, the respective  Newton system possesses a first integral
\[
I=z_jp_1-p_2.
\]
If in ~\eqref{eq:bez} all $n_i$ are greater than one, then the system can be
integrable.
We give such examples in the next section. 

\subsection{Infinitely many Darboux points}
\label{ssec:inf}
If
\begin{equation}
q_1F_2(q_1,q_2)=q_2F_1(q_1,q_2),
\label{eq:infinit}
\end{equation}
then force $\vF=(F_1,F_2)$ has infinitely many Darboux points. At each of these
points the Kovalevskaya exponents are $\Lambda_1=0$ and $\Lambda_2=k-1$.. Thus the
Morales-Ramis theory does not give any obstructions for  the
integrability.

It  is easy to show that condition~\eqref{eq:infinit} is satisfied if and only
if
\begin{equation}
\label{eq:infd}
F_1 = \sum_{i=0}^{k-2}f_i q_1^{k-1-i}q_2^i \mtext{and} F_2 = \sum_{i=0}^{k-2}f_i
q_1^{k-2-i}q_2^{i+1}.
\end{equation}
The Newton system with a force satisfying \eqref{eq:infinit}
possesses one first integral
\[
I=q_1p_2- q_2p_1.
\]

For $k=3$  and $k=4$ the Newton equations with the above form \eqref{eq:infd} of
forces are
 integrable with an
additional first integral  quadratic in the momenta
\[
\begin{split}
I_2 &= 3(f_0 p_1  + f_1 p_2)^2 +2 (f_0q_1 +f_1q_2)^3,\\
I_2&= 2( f_0p_1^2 + f_1p_1p_2 + f_2p_2^2) +(f_0q_1^2+f_1q_1q_2+f_2q_2^2)^2,
\end{split}
\]
respectively.  But for $k>4$, it seems that   the Newton equations with forces
given by~\eqref{eq:infd} are not 
integrable in a generic case, nevertheless,   they admit an additional
first integral  for some special values of $f_i$.   For example, if
\begin{equation}
 F_1 = q_1(f_0q_1+f_1q_2)^{k-2} \mtext{and} F_2=q_2(f_0q_1+f_1q_2)^{k-2},
\end{equation}
then the additional first integral has the following form
\begin{equation}
 I=k(f_0p_1+f_1p_2)^2+2(f_0q_1+f_1q_2)^k.
\end{equation} 
We do not know how to investigate  systematically the integrability of the
Newton equations in this case. It seems that we can apply only a
direct method. 

\begin{lemma}
 Assume that force $\vF=(F_1,F_2)$ admits infinitely many Darboux points
 and the
corresponding Newton equations have a polynomial first integral which is a
quadratic form with constant coefficients with respect to the momenta.  
Then these equations are equivalent to
an integrable Hamiltonian system with the radial potential 
\[
 H=\frac{1}{2}(p_1^2+p_2^2)+\frac{1}{k}(q_1^2+q_2^2)^{\frac{k}{2}},
\]
for $k$ even, or to the non-Hamiltonian integrable system
\[
 F_1=q_1q_2^{k-2},\qquad F_2=q_2^{k-1},\qquad
I_2=\frac{1}{2}p_2^2+\frac{1}{k}q_2^k,
\]
for any $k$.
\end{lemma}

\begin{proof}
 Without loss of the generality one can postulate the 
following form of the  first integral quadratic in the momenta
\begin{equation}
 I=\alpha p_1^2+p_2^2+\sum_{i=0}^kb_kq_1^{k-i}q_2^i.
\end{equation}
The case when the quadratic form of momenta is nonsingular 
corresponds (after
rescaling) to $\alpha=1$ and the case when this form is singular
corresponds to $\alpha=0$.
Condition $\dot I=0$ on solutions of \eqref{eq:infd} yields
\begin{equation}
 \begin{split}
 & (k-i)b_i-2\alpha f_i=0,\qquad i=0,\ldots k-2,\qquad b_{k-1}=0,\\
 & b_1=0, \qquad (i+2)b_{i+2}-2f_i=0,\qquad i=0,\ldots,k-2,
 \end{split}
\label{eq:recy}
\end{equation}
and the detailed analysis of solutions of this linear system 
yields the desired result.
\end{proof}

\section{Two degrees of freedom. Applications}
\label{sec:app}
In this section we consider also Newton equations with two degrees of freedom
but now we apply general theorems from the previous sections  to general 
homogeneous forces of low degrees. The first aim is to show that our general
results allow to perform the integrability analysis in a systematic and
algorithmic way.  Let us describe the main steps of this algorithm.

We fix $k>2$ and consider the Newton system with a homogeneous polynomial force
$\vF=(F_1,F_2)$ where $\deg F_1= \deg F_2=k-1$.  The problem is to  
give a complete list
of all $\vF$ for which the corresponding Newton system is integrable in the
Jacobi sense. As $\C$-linear space  $\C_{k-1}[\vq]$  of homogeneous 
polynomials
in two variables of degree $k-1$ has the dimension $k$, the problem 
depends on $2k$
parameters. We can reduce the dimension of the problem taking into 
account the
action of the four dimensional group $\mathrm{GL}(2,\C)$.

If for a given $k$  a complex number $\lambda$ belongs to an item in
table~\eqref{eq:tab} of Theorem~\ref{thm:MoRa}, then we write $\lambda\in\cN_k$.

In the  first step of the algorithm  we have to find for $1\leq l\leq k$ all
solutions of equation
\[ 
\sum_{i=1}^l\frac{1}{\Lambda_i} =-1,
\]
such that $\lambda_i=\Lambda_i+1\in\cN_k$ for $i=1,\ldots, l$. The set of 
such
solutions we denote as in~\cite{Maciejewski:05::b} by $\cI_{k,l}$. Let us 
remark
here that for small $k$ it can be determined easily with the help of a computer
algebra
program, however, for bigger $k$  it is a time demanding computational problem.

In the next step for each solution $(\Lambda_1,\ldots,\Lambda_l)$ obtained 
in
the previous step we have  to determine all non-equivalent forces $\vF$ with 
exactly $l$ simple
Darboux points, and such that at $i$-th point the non-trivial Kovalevskaya
exponent is $\Lambda_i$, for $i=1,\ldots,l$. For small $k$ it can be done
effectively, however, for bigger $k$ the problem is hard,
 because we have to solve a system of
nonlinear polynomial equations of high degrees. The forces selected in this way 
satisfy all
necessary integrability conditions of Theorem~\ref{thm:MoRa}  but of course
not all of them are integrable. To decide which of them are integrable,
 we have
to use a theorem giving the stronger necessary integrability conditions
 than those contained in~Theorem~\ref{thm:MoRa}. However, as far as we
know, there is no such a theorem.  In Appendix~B we present a possible approach
to this problem. It is a direct generalization of the method based 
on higher order variational
equations developed in~\cite{Morales:99::c,Morales:06::} for Hamilton equations.

The above steps do not finish the procedure. We have to investigate
separately all the cases when the force have lower than the maximal number of
Darboux points with a multiple Darboux point or does not have such a point at 
all, because for those cases we cannot apply Theorem~\ref{thm:MoRa}, 
but, nevertheless,
we can use Theorem~\ref{thm:less}. Moreover,
 in all considered cases we have to
distinguish all instances when the Newton system is Hamiltonian. 

Below, in a case when we cannot apply any theorem to check the integrability,  we use
several different algorithms for  finding a polynomial first integrals of a given
degree. We do not describe details of these computations and below we call
such a search   a direct method.

\subsection{Case $k=3$}
Here we investigate the Newton equations with a homogeneous
 force $\vF=(F_1,F_2)$
of degree two
\begin{equation}
\label{eq:k2}
F_1=\frac{a_{11}}{2}q_1^2+a_{12}q_1q_2+\frac{a_{13}}{2}q_2^2,\quad
F_2=\frac{a_{21}}{2}q_1^2+a_{22}q_1q_2+\frac{a_{23}}{2}q_2^2.
\end{equation}
Let us assume that $\vF$  possesses at least one Darboux point.  Then  it  can
be placed at the infinity, so 
 $F_1(0,q_2)=0$, and thus in \eqref{eq:k2} we put $a_{13}=0$.
Additionally, by a linear change of variables $q_1=a Q_1$ and $q_2=bQ_1+cQ_2$
where $ac\neq0$ one can obtain $a_{22}=0$.

As explained in Appendix~A, for $k=3$ the Newton equations are almost always 
 Hamiltonian.  After the above 
simplifications the Newton system is not Hamiltonian only when
\begin{align}
\label{eq:k3c1}
 a_{12}&=0 &&\mtext{and}&&a_{21}a_{23}\neq 0, &&\text{or}\\
\label{eq:k3c2}
 a_{21}&=0 &&\mtext{and} &&a_{12}(a_{23}-a_{12})\neq 0.&&
\end{align} 
Hamiltonian systems with homogeneous potentials of degree 3 were fully
analyzed in \cite{Maciejewski:04::g}, so  we consider here only
non-Hamiltonian
cases.  At first, let us assume that conditions~\eqref{eq:k3c1} 
are satisfied.
Then, as $a_{21}\neq 0 $ and $a_{23}\neq 0$, we can make a rescaling which
gives $a_{21}=-1 $ and $a_{23}=1 $. Hence we have to consider
the following one parameter family of Newton's equations
\begin{equation}
\begin{split}
\ddot q_1&=-\frac{1}{2}\alpha q_1^2, \qquad \alpha\in\C,  \\
\ddot q_2&=\frac{1}{2}q_1^2-\frac{1}{2}q_2^2.
\end{split}
\label{eq:ukladzik0}
\end{equation}
This system has one first integral
\begin{equation}
I=3p_1^2+\alpha q_1^3,
\label{eq:calka}
\end{equation}
and for the complete integrability one additional first integral is necessary.

For~\eqref{eq:ukladzik0} polynomials $g(z)$ and $h(z)$, see definitions~\eqref{eq:hg}, have
the forms
\[
g(z)=\frac{1}{2}(z^2+\alpha z-1),\qquad
h(z)=\frac{1}{2}\alpha.
\]
Thus 
\[
z_{1,2}=\frac{\alpha\pm\sqrt{\alpha^2+4}}{2},
\]
are two Darboux points provided $\alpha\neq 0$. If $\alpha\neq \pm 2\rmi$, then
$z_1\neq z_2$. 
The non-trivial Kovalevskaya exponents at these points
are
\begin{equation}
\label{eq:k3Li}
 \Lambda_{1,2}=\pm\sqrt{1+\frac{4}{\alpha^2}}.
\end{equation} 
Now, if system~\eqref{eq:ukladzik0} is integrable, then $\Lambda_{1}$ and
$\Lambda_{2}$ belong to table~\eqref{eq:tab} of Theorem~\ref{thm:MoRa}. It is
easy to notice that there are only two possibilities: either
$\Lambda_1=\Lambda_2=0$, or $\Lambda_1=-\Lambda_2=1$. 
But for $\Lambda_i$ given
by~\eqref{eq:k3Li} the second case is impossible. 
The first case gives $\alpha
=\pm 2\rmi$.

Summarizing, we have to investigate equations~\eqref{eq:ukladzik0} only when
$\alpha=\pm 2\rmi$ and $\alpha =0$.

Assume that $\alpha=0$. We introduce new coordinates 
\begin{equation*}
 Q_1 = q_1 +q_2 , \qquad Q_2 = q_1-q_2.
\end{equation*}
In these coordinates equations~\eqref{eq:ukladzik0} read
\begin{equation*}
 \ddot Q_1 = \frac{1}{2}Q_1Q_2, \qquad \ddot Q_2 = -\frac{1}{2}Q_1Q_2.
\end{equation*}
Applying Theorem~\ref{thm:less} to this system,
 we conclude that for $\alpha=0$
equations~\eqref{eq:ukladzik0} are not integrable in the Jacobi sense with
polynomial first integrals. 

If $\alpha=\pm 2\rmi$, then the considered force has a double Darboux point. 
At
this Darboux point the non-trivial Kovalevskaya exponent is $\Lambda=0$. Making
the following change of variables 
\begin{equation}
 Q_1= -\frac{\alpha}{4}q_1 + \frac{1}{2}q_2, \qquad Q_2=\frac{\alpha}{2}q_1,
\end{equation} 
we transform system~\eqref{eq:ukladzik0} for the prescribed values of $\alpha$
into the following one
\begin{equation}
 \ddot Q_1 = -Q_1(Q_1 + Q_2), \qquad \ddot Q_2 = - Q_2^2,
\end{equation} 
and now the double Darboux point is at infinity. 
 Let us consider a phase curve
$\Gamma_e$, $e\neq 0$,  corresponding to this Darboux point. It is given by 
\begin{equation*}
 e=\frac{1}{2}{\dot Q_2}^2 + \frac{1}{3}Q_2^3, \qquad Q_1=\dot Q_1= 0.
\end{equation*}
The particular solution corresponding to this phase curve is given by
$Q_2(t)=-6\wp(t,g_2,g_3)$, where invariants of the Weierstrass
 function are $g_2=0$ and $g_3= -e/18$. Variational equations
along this particular solution have the following form
\begin{equation}
 \ddot \xi_1 = 6\wp(t,g_2,g_3) \xi_1, \qquad \ddot \xi_2 = 12\wp(t,g_2,g_3)
\xi_2,
\end{equation} 
i.e., they are a direct product of two Lam\'e equation of the
Lam\'e-Hermite type. Assuming our Conjecture~\ref{con:hove} we can apply the
higher order 
variational equations for testing the integrability in the considered case.
Calculations show that a logarithmic term appears in the second order
variational
equations, so the system is not integrable. 

If condition~\eqref{eq:k3c2} is satisfied, then either we obtain a system
equivalent to~\eqref{eq:ukladzik0}, or to the following one 
\begin{equation}
 \label{eq:spec}
\ddot q_1 = -\lambda q_1 q_2, \qquad \ddot q_2 = -q_2^2, \qquad
\lambda\in\C^\star.
\end{equation} 
If $\lambda$ belongs to one of iteme in table~\eqref{eq:tab} (with $k=3$), then the necessary conditions of Theorem~\ref{thm:MoRa} are satisfied. It seems that for all these values the force  admits two or three independent polynomial first integrals.  
A family of forces with the above properties can be find for an arbitrary  $k>2$.  We discuss these families in the last subsection of this section. 

At the beginning, we assumed that the force admits at least one Darboux point.
Hence now, we consider all cases when there is no a Darboux point. By
Theorem~\ref{thm:g1} we can assume that the force has the form given
by~\eqref{eq:bez} with $k=3$. If in~\eqref{eq:bez} $p=2$, then $n_1=1$ or
$n_2=1$ and the system is not integrable. When $p=1$, $n_1=3$ the force has
the
following components
\begin{equation}
 F_1=(q_2-z_1q_1)(q_2-z_2q_1), \qquad F_2 = (q_2-z_1q_1)(q_2(2z_1-z_2)
-z_1^2q_1).
\end{equation}  
Let $
 Q_1 = q_1$ and $Q_2 = q_2 - z_1q_1$. Then in new variables the Newton equations
with the above force read
\begin{equation}
\label{eq:bezk3}
\ddot Q_1 =Q_2(Q_2 -\alpha Q_1), \qquad \ddot Q_2= -\alpha Q_2^2 , \qquad
\alpha\in\C,
\end{equation} 
and have one first integral
\begin{equation*}
 I = \frac{1}{2}P_2^2 +\frac{1}{3}\alpha Q_2^3, \qquad P_2=\dot Q_2.
\end{equation*}
If $\alpha=0$,  then system~\eqref{eq:bezk3} is a superintegrable Hamiltonian
system with the Hamilton function
\begin{equation*}
 H=P_1P_2 -\frac{1}{3}Q_2^3, 
\end{equation*}
and two additional first integrals
\begin{equation*}
 I_1 = P_2, \qquad I_2 = 4P_2(Q_1P_2-Q_2P_1)+Q_2^4.
\end{equation*}
If $\alpha\neq 0$, we can assume that $\alpha=1$. For this case, using a direct
 method, we checked that the system~\eqref{eq:bezk3} does not admit a second
polynomial first integral of degree lower than 12 with respect to the momenta. 
\subsection{Case $k=4$}
To shorten the exposition, we omit proofs and many details of calculations in
this
subsection.
At first, we recall from paper~\cite{Maciejewski:05::b} that  for $k=4$ set
$\cI_{4,4}$ 
of $\Lambda_i=\lambda_i-1$ with
$\lambda_i$ satisfying assumptions of Theorem~\ref{thm:MoRa} has elements
listed 
in Table~1, and $\cI_{4,l}=\emptyset$ for $2\leq l <4$.
\begin{table}[hb]
  \begin{center}
\begin{tabular}{|c|c|}
\hline 
element  & $\{\Lambda_1,\Lambda_2, \Lambda_3,\Lambda_4\}$ \\
\hline 
1    &  $\{-1,-1,2,2\}$ \\ 
2    & $\{-5/8,5,5,5\}$ \\
3    & $\{-5/8,2,20,20\}$ \\
4    & $\{-5/8,27/8,27/8,135\}$\\
5   & $\{-5/8,2,14,35\}$  \\
\hline
\end{tabular}
\end{center}
   \caption{Elements of set $\cI_{4,4}$.}
\end{table}

\subsubsection{Four Darboux points}
If we assume that a homogeneous force $\vF=(F_1,F_2)$ 
has four simple Darboux points, then, by a proper choice of coordinates, three of them
can be located arbitrarily on the 
Riemann sphere. Thus, we can assume
that one of them is at infinity, and two of them at $z=0$ and $z=1$,
respectively.  Under
these assumptions we have 

\begin{equation}
\begin{split}
F_1&=a_{11}q_1^3+a_{12}q_1^2q_2+a_{13}q_1q_2^2,\\
F_2&=a_{21}q_1^2q_2+a_{22}q_1q_2^2+a_{23}q_2^3,
\end{split}
\label{eq:k44}
\end{equation}
where 
\[
a_{11}+a_{12}+a_{13}=a_{21}+a_{22}+a_{23}=1,
\]
and
\[
 a_{11}a_{23}\neq0,\,\,\,\text{and}\,\,\, a_{11}\neq a_{21},\,\,\,\text{and}\,\,\, a_{13}\neq a_{23},\,\,\,\text{and}\,\,\, a_{11}-a_{21}\neq a_{13}-a_{23}.
\]
Now $z$-coordinates of Darboux points are
\begin{equation*}
z_0=0,\qquad z_1=1,\qquad z_2=\frac{a_{11}-a_{21}}{a_{13}-a_{23}},\qquad
z_3=\infty.
\end{equation*}
The respective non-trivial Kovalevskaya exponents at these points are the
following
\[
\Lambda_0=\frac{a_1}{a_{11}},\quad
\Lambda_1=a_3-a_1,
\quad\Lambda_2=\frac{(a_1-a_3)a_1a_3}{a_{11}a_3^2+a_{12}a_1a_3+a_{13}a_1^2},
\quad
\Lambda_3=-\frac{a_3}{a_{13}+a_3}, 
\]
where
\[
a_1=a_{21}-a_{11},\qquad a_2=a_{22}-a_{12},\qquad a_3=a_{23}-a_{13}.
\]

Now we can take the admissible values of $\Lambda_i$, $i=1,2,3,4$ from Table~1
and reconstruct $F_1$ and $F_2$.

For $(\Lambda_0,\Lambda_1,\Lambda_2,\Lambda_3)=(-1,-1,2,2)$ we obtain 
\begin{equation}
\begin{split}
F_1&=\frac{q_1}{5}[-(2a_2+1)q_1^2+5a_2q_1q_2+3(2-a_2)q_2^2],\\
F_2&=\frac{q_2^2}{5}[(a_2+3)q_1+(2-a_2)q_2],\qquad
a_2\not\in\left\{-\frac{1}{2},2\right\},
\end{split}
\label{eq:vv1}
\end{equation}
for $(\Lambda_0,\Lambda_1,\Lambda_2,\Lambda_3)=(5,5,-5/8,5)$:
\begin{equation}
\begin{split}
F_1&=\frac{q_1}{5}[(a_2-7)q_1^2+5a_2q_1q_2+6(2-a_2)q_2^2],\\
F_2&=\frac{q_2}{5}[6(a_2-7)q_1^2+5(9-a_2)q_1q_2+(2-a_2)q_2^2],\qquad 
a_2\not\in\{2,7\},
\end{split}
\label{eq:vv2}
\end{equation}
for $(\Lambda_0,\Lambda_1,\Lambda_2,\Lambda_3)=(2,20,-5/8,20)$:
\begin{equation}
\begin{split}
F_1&=\frac{q_1}{11}[10(a_2-22)q_1^2+11a_2q_1q_2+21(11-a_2)q_2^2],\\
F_2&=\frac{q_2}{11}[30(a_2-22)q_1^2+(660-29a_2)q_1q_2+(11-a_2)q_2^2],\qquad
 a_2\not\in\{11,22\},
\end{split}
\label{eq:vv3}
\end{equation}
for $(\Lambda_0,\Lambda_1,\Lambda_2,\Lambda_3)=(-5/8,27/8,27/8,135)$:
\begin{equation}
\begin{split}
F_1&=\frac{q_1}{220}[27(22-5a_2)q_1^2+220a_2q_1q_2-17(22+5a_2)q_2^2],\\
F_2&=\frac{q_2}{1760}[81(22-5a_2)q_1^2+410a_2q_1q_2-(22+5a_2)q_2^2],\qquad
a_2\not\in\left\{-\frac{22}{5},\frac{22}{5}\right\},
\end{split}
\label{eq:vv4}
\end{equation}
and for $(\Lambda_0,\Lambda_1,\Lambda_2,\Lambda_3)=(2,35,-5/8,14)$:
\begin{equation}
\begin{split}
F_1&=\frac{q_1}{16}[7(2a_2-77)q_1^2+16a_2q_1q_2+15(37-2a_2)q_2^2],\\
F_2&=\frac{q_2}{16}[21(2a_2-77)q_1^2+4(399-10a_2)q_1q_2+(37-2a_2)q_2^2],\quad
 a_2\not\in\left\{\frac{37}{2},\frac{77}{2}\right\}.
\end{split}
\label{eq:vv5}
\end{equation}
Permutations of allowable values of
$(\Lambda_0,\Lambda_1,\Lambda_2,\Lambda_3)$ give equivalent forces in all the
cases listed above. 

As it has already been mentioned, there is no forces satisfying conditions A.1
and A.2 formulated before 
Theorem~\ref{thm:rel} with a smaller number of simple Darboux points.

For forces~\eqref{eq:vv1}--\eqref{eq:vv5} the necessary integrability conditions
formulated in
Theorem~\ref{thm:MoRa} are satisfied. Thus, to check if
they are integrable, we applied the method of higher order variational
equations. Our calculations for forces~\eqref{eq:vv2}--\eqref{eq:vv5} show that
the only integrable cases correspond to
known Hamiltonian systems with homogeneous potentials. For a Newton equation with
 force \eqref{eq:vv1} the variational equations do not reduce to the product
of 
two Lam\'e equations and the application of higher order variational equations
is an open problem.

We also checked  whether the Newton equations with
forces~\eqref{eq:vv1}--\eqref{eq:vv5}
admit a single polynomial first integral for specific values of $a_2$. To this
end, we applied a direct method, and we looked for first integrals which have
degree lower than 8 with respect to the momenta. The results of these
calculations
are following. 

Force~\eqref{eq:vv1} admits a polynomial first integral only for
$a_2=-3$, but in this case the Newton equations are Hamiltonian and
integrable. 

Force~\eqref{eq:vv2} has  always a polynomial first integral
\[
\begin{split}
I_1&=q_1 (q_1-q_2) q_2 [(a_2-7)
    q_1-(a_2-2) q_2]^2-5 (a_2-7) q_2
    p_1^2-5 (a_2-2) q_1 p_2^2\\
    &+5 [(a_2-7)
    q_1+(a_2-2) q_2]p_1 p_2.
\end{split}
\]
Furthermore, for 
$a_2\in\{-3,9/2,12\}$ the system is Hamiltonian and integrable.  

For \eqref{eq:vv3}  if  $a_2=0$ the system is Hamiltonian  and integrable 
while for   $a_2=\frac{33}{31}(21\pm\sqrt{7})$  is Hamiltonian but 
non-integrable.  Except these Hamiltonian cases only for $a_2=22$ 
force~\eqref{eq:vv3} admits a polynomial first integral of the form 
\[
I=12 p_2^4+8 q_1 \left(q_2^7+4
    p_2^2 q_2^3\right)-4 p_2 (2 p_1+3
    p_2) q_2^4+3 q_2^8-8 q_1^2 q_2^6.
\]

Force \eqref{eq:vv4} admits polynomial first integrals only for $a_2=\{ 0,
\pm\frac{22}{25}\sqrt{\frac{4683}{43}}\}$ when the system is Hamiltonian and
non-integrable.

 Finally, force \eqref{eq:vv4} admits a polynomial first integral only for 
$a_2\in \{0, \frac{3}{22}(285\mp4\sqrt{86}) \}$ when the system is Hamiltonian
and non-integrable.

\subsubsection{Smaller number of simple Darboux points}
If the considered force admits fewer than four simple Darboux
points, than we have to investigate all the cases when condition A.2
formulated before Theorem~\ref{thm:rel} is not satisfied. It is easy to check 
that the case when  $k=4$, and the system possesses three
Darboux points is impossible.

Hence, we consider forces with two simple Darboux points. It can be shown that
in this case the force is given by 
\begin{equation}
\begin{split}
F_1&=q_1(q_1-q_2)[(\lambda_2-1)q_1+(\lambda_1-1)\lambda_2q_2],\\
F_2&=q_2(q_1-q_2)[(\lambda_1-1)q_2+(\lambda_2-1)\lambda_1q_1].
\end{split}
\label{eq:hrrrum}
\end{equation}
provided that $\lambda_1\neq 1$ and $\lambda_2\neq 1$. If $\lambda_1=1$ or
$\lambda_2 =1$, then the above force has infinitely many Darboux points and, as
we showed in Section~\ref{ssec:inf}, the corresponding Newton system is
integrable.
The above force satisfies the necessary conditions of Theorem~\ref{thm:MoRa} iff
$\lambda_1$ and $\lambda_2$ take values from table~\eqref{eq:tab}.
Let us notice that to derive~\eqref{eq:hrrrum} we have to use a generalized
relation~\eqref{eq:genrel} as described in Remark~\ref{rem:genrel}. 

Components of force~\eqref{eq:hrrrum} have a common factor $q_1-q_2$, so we can
use~Theorem~\ref{thm:less}. Applying Proposition~\ref{prop:rot} we easily
transform the force to the form $F_i=q_2\widetilde F_i$, where $\widetilde F_i$
are
homogeneous polynomials of degree 2 for $i=1,2$. It appears that $\deg_{q_1}
\widetilde F_2 = 2$ iff $\lambda_1\neq\lambda_2$. Hence, by
Theorem~\ref{thm:less}, the Newton system with force~\eqref{eq:hrrrum} is not
integrable if $\lambda_1\neq\lambda_2$. 

If $\lambda = \lambda_1=\lambda_2\neq 1$ is given by an item from
table~\eqref{eq:tab}, then, in some cases, we were able to find one first
integral. 
For example, if $\lambda=6$, then the first integral has the form 
\[
 I=q_2p_1^2+q_1p_2^2-(q_1+q_2)p_1p_2+5q_1q_2(q_2-q_1)^3.
\]
 Also when $\lambda_1\neq\lambda_2$ we can find
one first integral. For example, for $\lambda_1=3/8$ and $\lambda_2=6$ 
the first integral reads
\[
I=p_2(q_1p_2-q_2p_1)-\frac{5}{8}q_1q_2^2(q_1-q_2)^2.
\]

Now we consider the case with only one Darboux point. It can be shown that in
this case the force is given by
\begin{align}
 \label{eq:1d1}
F_1 &= \alpha q_1q_2(q_1+q_2), && F_2=q_2(q_1+q_2)[(1-\alpha)q_1+q_2],&&
\alpha\neq 1,&&\text{or}\\
 \label{eq:1d2}
F_1&= \alpha q_1^2q_2,&& F_2=q_2^2(q_2+ \alpha q_1), && &&\text{or}\\
 \label{eq:1d3}
F_1 &= \lambda q_1 q_2^2, && F_2 = q_2^3.&& &&
\end{align}  
By Theorem~\ref{thm:less}, the Newton system with force~\eqref{eq:1d1} is not
integrable. For $\alpha=0$ the Newton equations with force~\eqref{eq:1d2} are
Hamiltonian and integrable, while for  $\alpha\neq 0$ they are also
Hamiltonian,  however nonintegrable, as it was proved in
\cite{Maciejewski:05::b}.  
The case of force~\eqref{eq:1d3} is discussed in the last
subsection of this section.

Assuming that Conjecture~\ref{con:hove} is valid,  and  applying the 
 higher order variational equations  method similarly  as it was done for
Hamiltonian systems in
\cite{Maciejewski:04::g, Maciejewski:05::b}, it is possible to prove the
following.
\begin{lemma}
 The only integrable Newton systems with $k=4$ possessing a multiple Darboux
point are equivalent to the
following one
\[
\ddot q_1=-q_1(a_{11}q_1^2+q_2^2),\qquad
\ddot q_2=-q_2(a_{11}q_1^2+q_2^2),
\]
with first integrals
\[
 I_1=q_1p_2-q_2p_1,\qquad
I_2=\frac{1}{2}(a_{11}p_1^2+p_2^2)+\frac{1}{4}(a_{11}q_1^2+q_2^2)^2.
\]
\end{lemma}
Finally, let us consider the integrability problem for forces without a Darboux
point. 
Then by Theorem~\ref{thm:g1} the only forces that can admit two first integrals
are  either of the following:
\begin{align}
\label{eq:k4bd1}
 F_1& = q_2[q_2(q_1+\alpha q_2)+q_1^2],&& F_2=q_2^2(q_1+q_2),&&\text{or}\\
\label{eq:k4bd2}
 F_1 &= 0, && F_2 =-q_1q_2^2,&&\text{or}\\
\label{eq:k4bd3}
  F_1 &= q_1^2q_2, &&  F_2=-\alpha q_1 q_2^2,&&\text{or}\\
\label{eq:k4bd4}
  F_1 &= 0, &&  F_2=q_1^3,&&\text{or}\\
\label{eq:k4bd5}
  F_1 &= q_2^2(q_1+q_2), &&  F_2=q_2^3.&&
\end{align} 
The  force field \eqref{eq:k4bd1}  for any $\alpha$ possesses a first integral
that is a nondegenerate quadratic form with respect to the momenta with constant
coefficients, thus it is always Hamiltonian, however we found the second
functionally independent first integral only for $\alpha=0$. Calculations were
made up to degree 8 in the momenta. 
The force with components \eqref{eq:k4bd2} has one first integral $I=p_1$ and
there is no the second polynomial first integral  up to degree 9 in the momenta.
Newton system governed by \eqref{eq:k4bd4} is integrable and Hamiltonian. The
last force \eqref{eq:k4bd5} is non-Hamiltonian with one first integral
$I=2p_2^2+q_2^4$ and the direct method does not yield the second polynomial
first integral up to degree 10 in the momenta. 

The most puzzling are forces \eqref{eq:k4bd3}.   The application of the direct method up to degree 7 in the momenta  yields one integrable Hamiltonian case
for $\alpha=-1$, the another one integrable  non-Hamiltonian for
$\alpha=-11/6$ with  two first integrals which are both of degree four in the momenta
\[
 \begin{split}
 & I_1=24p_1p_2^3+3q_2^2(4q_1^2p_2^2 + 12q_1q_2p_1p_2 -
3q_2^2p_1^2)+16q_1^3q_2^5,\\
&I_2=162p_1^3(q_1p_2  -q_2p_1)-9q_1^3(4q_1^2p_2^2 - 20q_1q_2p_1p_2 +
13q_2^2p_1^2)+
16q_1^6q_2^3.
 \end{split}
\]
Moreover we found that for some rational and irrational values of $\alpha$ one polynomial first integral exists. This is the case for 
 $\alpha\in\{0,-7/2,-4,-5,-10,-11,(-38+21\sqrt{3})/11,-187/7\}$.  For these  values of $\alpha$  first integrals are quite complicated as it is shown by some examples below:
\begin{align*}
\alpha&=-\frac{7}{2},&&I=2p_1(q_1p_2-q_2p_1)+q_1^3q_2^2,\\
\alpha&=-4,&&I=4p_1^3-q_1^3(q_1p_2-4q_2p_1),\\
\alpha&=-5,&&I=6p_1^3p_2+q_1^2(q_1^2p_2^2 - 4q_1q_2p_1p_2 +
15q_2^2p_1^2)+2q_1^5q_2^3,\\
\alpha&=\frac{-38+21\sqrt{3}}{11},&&I=176(38 + 21\sqrt{3})p_1^3p_2^3-33[(39 +
23\sqrt{3})q_1^4p_2^4 - 4(39 + 23\sqrt{3})q_1^3q_2p_1p_2^3 \\
&&&-10(7 + 3\sqrt{3})q_1^2q_2^2p_1^2p_2^2 + 4(-3 + 5\sqrt{3})q_1q_2^3p_1^3p_2 +
(3 - 5\sqrt{3})q_2^4p_1^4]\\
&&&+264q_1^3q_2^3[3(2 + \sqrt{3})q_1^2p_2^2 - 2q_1q_2p_1p_2 -3(-2 +
\sqrt{3})q_2^2p_1^2]\\
&&&-32(-7 + 3\sqrt{3})q_1^6q_2^6.
\end{align*}

\subsection{Case $k>4$}

We performed an analysis of the case with $k=5$ in the same way as for the case
with $k=4$ in the
previous section.   It does not give any intersting examples of Newton's equations
except for a case when the force admits only one simple Darboux point (see the last subsection) or when they do not admit any Darboux point.  It should be mentioned that the bigger number of parameters make the analysis more complicated. 
 
Below we describe a few general facts which are valid for arbitrary $k>5$. 
Let us consider a case when the force has exactly $k$ Darboux points.
Then for each $k$ set $\cI_{k,k}$ contains  two elements
\begin{equation}
 \cI_{k,k}^{(1)}=\{-1,-1,\underbrace{k-2,\ldots,k-2}_{k-2\
\text{times}}\},\qquad
\cI_{k,k}^{(2)}=\left\{-\frac{k+1}{2k},\underbrace{k+1,\ldots,k+1}_{k-1\
\text{times}}\right\}.
\end{equation} 
Let us note that for $k=14, 17, 19, 26, \ldots$ this set is bigger, see
\cite{mp:06::h}.
We show the following fact.
\begin{lemma}
Let  $\vF=(F_1,F_2)$ be a homogeneous force possesing $k$ Darboux
points, where $k=1+\deg F_i$. Assume that $(\Lambda_1,\ldots,
\Lambda_k)=\cI_{k,k}^{(2)}$, where $\Lambda_i$ is the non-trivial
Kovalevskaya exponent of $\vF$ at $i$-th Darboux point. Then $\vF$ is
equivalent to the following force
\begin{equation}
\begin{split}
F_1&=\sum_{i=1}^{k-1}ig_iq_1^{k-i}q_2^{i-1},\\
F_2&=\sum_{i=0}^{k-1}(k+1+i)g_iq_1^{k-1-i}q_2^{i}.
\end{split}
\label{eq:2fama}
\end{equation}
\end{lemma}
\begin{proof}
We can assume that one Darboux point is at infinity. The polynomial $g(z)$ has
degree $k-1$ and the degree of polynomial $h(z)$ is smaller than $k-1$, see
formulae~\eqref{eq:hg}. We put
\begin{equation}
\label{eq:geg}
 g(z)=\sum_{i=0}^{k-1}g_iz^i=g_{k-1}\prod_{i=1}^{k-1}(z-z_i).
\end{equation} 
Then
\[
\frac{h(z)}{g(z)}=\sum_{i=1}^{k-1}\frac{g'(z_i)}{h(z_i)}\frac{ 1 }{z-z_i}
=\sum_{i=1}^{k-1}\frac{1}{\Lambda_i}\frac{1}{z-z_i}.
\]
Since $\Lambda_1=\cdots=\Lambda_{k-1}=k+1$, we have
\[
h(z)=\frac{1}{k+1}\sum_{i=1}^{k-1}\left(\frac{1}{z-z_i}\right)g(z)=\frac{1}{k+1}
g'(z).
\]
As result 
\[
f_1(z)=\frac{1}{k+1}g'(z),\qquad f_2(z)=g(z)+zh(z)=g(z)+ \frac{1}{k+1}zg'(z).
\]
Direct substitution of the explicit form  of $g(z)$ yields
\[
\begin{split}
f_1&=\frac{1}{k+1}\sum_{i=1}^{k-1}ig_iz^{i-1},\\
f_2&=\frac{1}{k+1}\sum_{i=0}^{k-1}(k+1+i)g_iz^{i},
\end{split}
\]
and homogenization of $(f_1,f_2)$ and multiplication by $k+1$ finishes the
proof.
\end{proof}

One can check directly that
\[
I=(k+1)p_1(q_1p_2-q_2p_1)+\sum_{i=1}^kg_{i-1}q_1^{k+2-i}q_2^{i-1},
\]
is a first integral of Newton's equations with force 
\eqref{eq:2fama}. For this force the necessary integrability conditions of
Theorem~\ref{thm:MoRa}
Morales-Ramis theory are satisfied.  Thus we can only apply a direct method
for searching the second first integral. Here the problem is not trivial as
force~\eqref{eq:2fama} depends on many parameters. We have not been able to make
a
general search for first integrals of degree higher than 2 with respect to
momenta. In all  cases when a first integral was found, the system appeared to
be  Hamiltonian or trivially
integrable. 

\begin{lemma}
Let  $\vF=(F_1,F_2)$ be a homogeneous force possesing $k$ Darboux
points, where $k=1+\deg F_i$. Assume that $(\Lambda_1,\ldots,
\Lambda_k)=\cI_{k,k}^{(1)}$, where $\Lambda_i$ is the non-trivial
Kovalevskaya exponent of $\vF$ at $i$-th Darboux point. Then $\vF$ is
equivalent to the following force
\begin{equation}
 F_1=\sum_{i=1}^{k-1}(i+1-k)g_iq_1^{k-i}q_2^{i-1},\qquad
F_2=\sum_{i=1}^{k-1}(i-1)g_iq_1^{k-i-1}q_2^i.
\label{eq:1fama}
\end{equation}
\end{lemma}
\begin{proof}
We can assume that one Darboux point is 
 at the infinity  and  for it $\Lambda_0=-1$. The other with 
$\Lambda_{k-1}=\Lambda_-=-1$  can be locate at  $z_{k-1}=0$.  
For Darboux points $z_1,\ldots,z_{k-2}$ we have
$\Lambda_1=\cdots=\Lambda_{k-2}=\Lambda=k-2$. Thus 
\[
\frac{h}{g}=\frac{1}{\Lambda}\sum_{i=1}^{k-2}\frac{1}{z-z_i}+\frac{1}{\Lambda_{-
}}\frac{1}{z-z_{k-1}}=
\frac{1}{\Lambda}\sum_{i=1}^{k-1}\frac{1}{z-z_i}+\left(\frac{1}{\Lambda_{-}}
-\frac{1}{\Lambda}\right)\frac{1}{z-z_{k-1}},
\]
or, in other words
\[
h=\frac{1}{\Lambda}g'+\left(\frac{1}{\Lambda_{-}}-\frac{1}{\Lambda}\right)\frac{
g}{z-z_{k-1}}=\frac{g'(z)}{k-2}-\frac{(k-1)g(z)}{(k-2)(z-z_{k-1})}.
\]
We recall that   $f_1=h$ and $f_2=g+zh$. Thus, putting $z_{k-1}=0$
and
multiplying by $k-2$, we obtain
\[
f_1(z)=g'(z)-\frac{(k-1)g(z)}{z},\qquad 
f_2(z)=zg'(z)-g(z).
\]
Substitution of \eqref{eq:geg} into the above expressions and homogenization
gives~\eqref{eq:1fama}.
\end{proof}
For  force~\eqref{eq:1fama} the necessary integrability conditions of
Theorem~\ref{thm:MoRa} are satisfied. Hence, as in the previous case, we can
only
apply a direct search for two first integrals, however conclusions from such an
investigation are the same as for force~\eqref{eq:2fama}. 

\subsection{An integrable family of Newton equations}
\label{ssec:intfam}
Here we consider a force given by
\begin{equation}
 \label{eq:ifam}
 F_1 = \lambda q_1 q_2^{k-2}, \qquad F_2 =q_2^{k-1}, \qquad \lambda\in
\C^\star,\quad k>2.
\end{equation} 
This force has one Darboux point at infinity, and it satisfies the necessary
conditions for integrability of Theorem~\ref{thm:MoRa} provided $\lambda$
belongs to
an item of table~\ref{eq:tab}. It appears that for all these values of
$\lambda$, the Newton equations are really integrable. Moreover, this
integrability is highly non-trivial. 

Force~\eqref{eq:ifam} admits one obvious first integral
\begin{equation}
 \label{eq:I1}
I_1 = \frac{1}{2}p_2^2 +\frac{1}{k}q_2^2. 
\end{equation} 
Assume that 
\begin{equation}
 \label{eq:l1}
\lambda = p + \frac{k}{2}p(p-1), \qquad  p\in\Z^\star.
\end{equation} 
Then for each $k$ and $p$ there exists an additional first integral of degree
$p$ with respect to the momenta for $p>0$, for $p<0$ this degree is $1-p$. For
example, if $k=3$ and $p=6$ this first integral reads
\begin{equation*}
 I_2=28p_2^5(q_1p_2-q_2p_1)-70q_2^3p_2^3(6q_1p_2-q_2p_1)+15q_2^6p_2(21q_1p_2-q_2p_1)-15q_1q_2^{9},
\end{equation*}
while for $k=6$ and $p=-5$ it takes the form
\begin{equation*}
 I_2 =11p_1p_2^5+11q_2^5p_2^3(17q_1p_2-2q_2p_1)-3q_2^{11}
p_2(34q_1p_2-q_2p_1)+3q_1q_2^{17}.
\end{equation*}
We checked the above claim finding explicit forms of integrals for $k<10$ and
$\abs{p}<10$. Let us note that for  natural Hamiltonian systems with a
homogeneous potential in all known integrable cases the additional first
integral is of the degree smaller than 5 with respect to the momenta. 

If we assume that 
\begin{equation}
 \label{eq:l11}
\lambda = \frac{1}{2}\left[ \frac{k-1}{k} +kp(p+1)\right] , \qquad 
p\in\N\cup\{0\},
\end{equation} 
the force admits two additional first integrals, so the corresponding Newton
equations are superintegrable with three polynomial first integrals. Degrees of
two additional integrals with respect to the momenta depend on $p$ and are
unbounded with respect to $p$. For example, for $k=3$ and $p=0$ the additional
first integrals read
\begin{equation*}
  I_2=6p_1(q_1p_2-q_2p_1)+q_1^2q_2^2,\quad I_3=6p_1^3-q_1^2(q_1p_2-3q_2p_1),
\end{equation*} 
while for $k=4$ and $p=1$ they have the following forms
\begin{equation*}
 \begin{split} 
I_2&=240p_1p_2^2(q_1p_2-q_2p_1)+2q_2^3(175q_1^2p_2^2-260q_1q_2p_1p_2+4q_2^2p_1^2
) -225q_1^2q_2^7,\\
I_3&=10368p_1^4p_2^2+16(625q_1^4p_2^4-2500q_1^3q_2p_1p_2^3+3750q_1^2q_2^2p_1^
2 p_2^2+1280q_1q_2^3p_1^3p_2+4q_2^4p_1^4)\\
&-1350q_1^2q_2^4(25q_1^2p_2^2-80q_1q_2p_1p_2-8q_2^2p_1^2)+50625q_1^4q_2^8.
 \end{split}
\end{equation*}
We checked the above claim as the previous one, finding explicit forms of first
integrals. In all cases we checked that three first integrals $I_1$, $I_2$
and $I_3$ are functionally independent. 

For $k>5$ the considered above two choices for $\lambda$ are the only admitted
by Theorem~\ref{thm:MoRa}. For $k\leq5$ we have additional admissible choices
for $\lambda$. For example, for $k=3$ the four additional families depending on
integer parameter $p$ appear.
 Below we show first integrals for a few elements of these families:
for  $\lambda=-\frac{1}{24} +\frac{1}{6}(1+3p)^2$
 \begin{align*}
  p&=0,&&I_2=192p_1^4-16q_1^2p_1(q_1p_2-3q_2p_1)-q_1^4q_2^2,\\
&&&I_3=192p_1^3(q_1p_2-q_2p_1)+2q_1^2(q_1^2p_2^2 - 4q_1q_2p_1p_2 +
24q_2^2p_1^2)+
q_1^4q_2^3,\\
p&=1,&&I_2=5184p_1^4p_2^2-16p_1(343q_1^3p_2^3 - 1029q_1^2q_2p_1p_2^2
-672q_1q_2^2p_1^2p_2 + 8q_2^3p_1^3)\\
&&&-147q_1^2q_2^2(49q_1^2p_2^2 - 224q_1q_2p_1p_2
-32q_2^2p_1^2)+14406q_1^4q_2^5,\\
&&&I_3=36288p_1^3p_2^4(q_1p_2-q_2p_1)+14p_2^2(343q_1^4p_2^4 -
1372q_1^3q_2p_1p_2^3 + 12264q_1^2q_2^2p_1^2p_2^2\\
&&& -
12928q_1q_2^3p_1^3p_2 + 640q_2^4p_1^4)
-q_2^3(7203q_1^4p_2^4 - 208544q_1^3q_2p_1p_2^3 +
272832q_1^2q_2^2p_1^2p_2^2 \\
&&&- 16128q_1q_2^3p_1^3p_2 - 256q_2^4p_1^4)+588q_1^2q_2^6(147q_1^2p_2^2 -
280q_1q_2p_1p_2 +
16q_2^2p_1^2)-28812q_1^4q_2^9,
\end{align*}
for $\lambda=-\frac{1}{24} +\frac{3}{32}(1+4p)^2$
\begin{align*}
 p&=0,&&I_2=294912p_1^6+15360q_1^2p_1^3(-q_1p_2 + 3q_2p_1)-16q_1^4(q_1^2p_2^2 -
6q_1q_2p_1p_2 + 90q_2^2p_1^2)-9q_1^6q_2^3,\\
&&&I_3=28311552p_1^7(q_1p_2-q_2p_1)+516096q_1^2p_1^4(q_1^2p_2^2 - 4q_1q_2p_1p_2
+ 16q_2^2p_1^2\\
&&&+1536q_1^4p_1(-q_1^3p_2^3 + 7q_1^2q_2p_1p_2^2 +14q_1q_2^2p_1^2p_2 +
84q_2^3p_1^3)-8q_1^6q_2^2(5q_1^2p_2^2 + 132q_1q_2p_1p_2\\
&&& -1008q_2^2p_1^2)-27q_1^8q_2^5.
\end{align*}
 For $\lambda=-\frac{1}{24}+\frac{3}{50}(1+5p)^2$ and for
$\lambda=-\frac{1}{24}+\frac{3}{50}(2+5p)^2$ 
expressions for $I_2$ and $I_3$  are more complicated and we do not write them
here.

Our calculations show  that for additional families of $\lambda$'s there always
exist
two additional functionally independent first integrals.

The investigations shortly presented above allowed us to formulate
Conjecture~\ref{con:ifam}.
Without doubt, the integrability analysis of
equations~\eqref{eq:nifam} presented above is not satisfactory, and  a separate
paper
\cite{Maciejewski:07::a} will be devoted to more involved
investigations.

\renewcommand{\thesection}{\Alph{section}}
\setcounter{section}{1}
\setcounter{equation}{0}
\section*{Appendix A}
Here we characterize in two ways forces that are potential in the sense of
Definition~\ref{def:pot}. 

\begin{lemma}
 \label{lem:ham}
Assume that Newton's equations
\begin{equation}
\label{eq:nn}
 \dot \vq = \vp, \qquad \dot \vp = -\vF(\vq), \qquad \vq, \vp\in\C^n,
\end{equation} 
 admit a first integral of the form 
\begin{equation}
 \label{eq:I}
I= \frac{1}{2}\vp^TK\vp +W(\vq),
\end{equation}
where $K\in\mathrm{GL}(n,\C)$, $K=K^T$.  Then force $\vF$ is potential and 
equations~\eqref{eq:nn}  are  Hamiltonian with respect to a symplectic
structure
given by the following matrix
\begin{equation}
 J=\begin{bmatrix}
      \vzero & K^{-1}\\
     -K^{-1} & \vzero
     \end{bmatrix},
\end{equation}
and $I$ becomes the Hamiltonian function. 
\end{lemma}
\begin{proof}
 Since $I$ is a first integral of equations~\eqref{eq:nn}, we have
\begin{equation*}
 \vp^T\pder{W}{\vq} +\vp^TK\vF=0,
\end{equation*}
so
\begin{equation*}
 \vF = K^{-1}\pder{W}{\vq} = K^{-1}\pder{I}{\vq}.
\end{equation*}
Hence
\begin{equation*}
 \begin{bmatrix}
 \vp\\
-\vF
\end{bmatrix}
=\begin{bmatrix}
      \vzero & K^{-1}\\
     -K^{-1} & \vzero
     \end{bmatrix}
\begin{bmatrix}
 \pder{I}{\vq}\\[1em]
\pder{I}{\vp}
\end{bmatrix}.
\end{equation*}
In this way we proved the second claim of our lemma. To prove the first one, let
us notice that for every non-singular matrix $K$ there exists a matrix
$A\in\mathrm{GL}(n,\C)$ such that $K= AA^T$.  Hence, in new variables
\begin{equation}
 \vQ =A\vq, \quad \vP=A\vp, \mtext{where} AA^T =K,
\end{equation} 
equations~\eqref{eq:nn} read
\begin{equation}
 \dot \vQ=\vP, \qquad \dot\vP = -\nabla V(\vQ), \quad  V(\vQ) := W(A^{-1}\vQ),
\end{equation} 
so they are canonical Hamilton's equations with
\begin{equation}
 H:=I(A^{-1}\vQ,A^{-1}\vP)=\frac{1}{2}\vP^T\vP + V(\vQ),
\end{equation} 
as the  Hamiltonian function.
\end{proof}
The second characterization of potential forces is the following.
If a force $\vF(\vq)$ is potential, then, by Definition~\ref{def:pot}, there
exists a non-singular matrix $A$ such that 
\[
\vF_A(\vQ)=A^{-1}\vF(A\vQ)=\frac{\partial V(Q)}{\partial Q},
\]
for some scalar function $V$.
Hence, putting $\vq=A\vQ$ we obtain 
\begin{equation}
K\vF(\vq)=\frac{\partial V(\vq)}{\partial \vq},\qquad K^{-1}= A^{T}A. 
\label{eq:warun0}
\end{equation}
This shows that $\vF$ is a potential iff there exists a non-singular matrix $K$
such that differential form 
\begin{equation}
w=\langle K\vF,\mathrm{d}\vq\rangle :=\sum_{i=1}^n \sum_{j=1}^n K_{ij}F_j \rmd
q_i,
\label{eq:warun}
\end{equation}
is exact. 

The condition $\rmd w =0$ is equivalent to the following set of equations
\begin{equation}
\sum_{i=1}^n\left(K_{pi}\frac{\partial F_i}{\partial q_r}-K_{ri}\frac{\partial
F_i}{\partial q_p}\right)=0, \qquad 1\leq p<r\leq n.
\label{eq:waham}
\end{equation}
For a given force in the above equations the elements of matrix are the unknowns
$K=[K_{ij}]$ with $i\leq j$. Since we consider the polynomial force and $\deg F_i=k-1$, all
$n(n-1)/2$ equations~\eqref{eq:waham} are polynomial and homogeneous of degree
$k-2$, Hence, they give rise to a system of homogeneous linear
equations for unknowns $K_{ij}$.

As an example we consider the case $n=2$ and $k=3$. We put 
\begin{equation}
\label{eq:kk2}
F_1=\frac{a_{11}}{2}q_1^2+a_{12}q_1q_2+\frac{a_{13}}{2}q_2^2,\quad
F_2=\frac{a_{21}}{2}q_1^2+a_{22}q_1q_2+\frac{a_{23}}{2}q_2^2.
\end{equation}
Then equations \eqref{eq:waham} are equivalent to the following linear
system
\begin{equation}
L \vk=0,\qquad
L=\begin{pmatrix}
a_{13}&a_{23}-a_{12}&-a_{22}\\
a_{12}&a_{22}-a_{11}&-a_{21}
    \end{pmatrix}
\label{eq:syso}
\end{equation}
and $\vk=[K_{11},K_{12},K_{22}]^T$.
This system has always a non-zero solution. If one of $2\times 2$ minors $L_i$
of matrix $L$ 
\[
L_1=a_{12}(a_{12}-a_{23})+a_{13}(a_{22}-a_{11}),\,\,  L_2=a
_{21}(a_{12}-a_{23})+a_{22}(a_{22}-a_{11}),\,\,
L_3=a_{12}a_{22}-a_{13}a_{21},
\]
is non-zero, then it has the follwing solution 
\[
\begin{split}
K_{11}&=a_{21}(a_{12}-a_{23})+a_{22}(a_{22}-a_{11}),\qquad
K_{12}=a_{13}a_{21}-a_{12}a_{22},\\
K_{22}&=a_{13}(a_{22}-a_{11})+a_{12}(a_{12}-a_{23}),
\end{split}
\]
and matrix $K$ given by this solution is not singular iff
\begin{equation}
\label{eq:kond1}
\begin{split}
&a_{12}^2[a_{12}a_{21}-a_{11}a_{22}-2a_{21}a_{23}]+
a_{12}[(a_{11}-a_{22})a_{22}a_{23}-a_{13}a_{21}(a_{11}-3a_{22})+a_{21}a_{23}^2]
\\
&+a_{13}[(a_{11}-a_{22})((a_{11}-a_{22})a_{22}+a_{21}a_{23})-a_{13}a_{21}^2]
\neq 0.
\end{split}
\end{equation}

For values of $a_{ij}$ for which all minors $L_i$ vanish 
one can find a two-parameter family of solutions of \eqref{eq:syso} and always
obtain a non-singular matrix $K$.

For $n=2$ and $k>3$, as well as, for $n>2$ equations~\eqref{eq:waham} give rise to an overdetermined
system of linear homogeneous equations, so a non-zero solution exists only
if the rank of matrix $L$ is not maximal. As entries of $L$ depend on
coefficients of polynomials $F_1$ and $F_2$, this gives very
restrictive conditions for a force $\vF=(F_1,F_2)$ to be potential. 
\stepcounter{section}
\section*{Appendix B}
Assume that we know a particular solution $\vvarphi(t)$ of a holomorphic system 
\begin{equation}
 \label{eq:hds}
\Dt \vx = \vv(\vx), \qquad \vx\in\C, \quad t\in\C.
\end{equation} 
Then appart from variational equations 
\begin{equation}
 \Dt \vxi = A(t)\vxi, \qquad A(t)= \pder{\vv}{\vx}(\vvarphi(t)),
\end{equation} 
one can use higher order variational equations. The idea is  following. Let us
put
\[
\vx=\vvarphi(t)+\varepsilon\vxi^{(1)}+\varepsilon^2\vxi^{(2)}+\cdots+
\varepsilon^k\vxi^{(k)}+\cdots,
\]
where $\varepsilon$ is a formal small parameter. Inserting the above
expansion into equation \eqref{eq:hds} and comparing terms of the
same order with respect to $\varepsilon$, we obtain the following chain
of linear non-homogeneous equations
\begin{equation}
\label{eq:vek}
\Dt\vxi^{(k)} = A(t)\vxi^{(k)} + \vf_k(\vxi^{(1)},
\ldots, \vxi^{(k-1)}), \qquad k=1,2, \ldots,
\end{equation}
where
$\vf_1\equiv 0$. For a given $k$ equation \eqref{eq:vek} is called the 
$k$-th order variational equation.
We denote by $X(t)$ the fundamental matrix of the homogeneous system, i.e.,
$n\times n$ matrix satisfying
\[
\Dt X = A(t) X, \qquad X(0) = E, 
\]
where $E$ is the identity matrix. 
Then solutions of $k$-th order variational equations for $k>1$ are given by
\begin{equation}
\label{eq:xik}
\vxi^{(k)}(t) = X(t)\vc(t),
\end{equation}
where $\vc(t)$ is a solution of
\begin{equation}
\label{eq:ct}
\Dt \vc = X^{-1}(t) \vf_k.
\end{equation}

There exists an
appropriate framework allowing to define the differential Galois group
of the $k$-th order variational equation, for details
see~\cite{Morales:99::c,Morales:06::}.  For Hamiltonian systems the following theorem was
proved in~\cite{Morales:06::}.

\begin{theorem}
\label{thm:ho}
  Assume that a Hamiltonian system is meromorphically integrable in
  the Liouville sense in a neighborhood of a phase curve 
  $\Gamma$ which is not an equilibrium point.  Then the identity component of
the differential Galois
  group of the $k$-th order variational equations is
  Abelian for any $k\in\N$.
\end{theorem}
The main problem with an application of the above theorem lies in the fact that
it is very difficult to investigate the differential Galois group of higher
order variational equations. However, if the first order variational equations
are a direct product of Lam\'e equations, then there is a local criterion which
allows to decide whethere this group is Abelian, for
details and practical applications
see~\cite{Morales:99::c,Morales:06::,Maciejewski:04::g,Maciejewski:05::b}. 

Let us note that the main construction presented in~\cite{Morales:06::} is
valid also for    non-Hamiltonian systems. Thus, we can also use it for the
Newton's equations. Unfortunately, there are many theoretical problems in
extending
Theorem~\ref{thm:ho} in such way that it should apply to the Newton equations.
Nevertheless, there are many facts strongly suggesting that the following
conjecture is true.
\begin{conjecture}
\label{con:hove}
 Assume that a Newton system of the form~\eqref{eq:newton1} is meromorphically
integrable in
  the Jacobi sense in a neighborhood of a phase curve 
  $\Gamma$ corresponding to a Darboux point.  Then the identity component of
the differential Galois
  group of the $k$-th order variational equations is
  Abelian for any $k\in\N$.
\end{conjecture}

\section*{Acknowledgments}
The author is very grateful to Andrzej Maciejewski for many helpful comments 
and suggestions concerning improvements and simplifications of some results.
The author also thanks Alain Albouy for his fruitful comments about
quasi-Lagrangian systems and sending Halphen's papers.
This research has been partially supported by  the European Community project
GIFT  (NEST-Adventure Project no. 5006) and by projet de l'Agence National de la
Recherche
"Int\'egrabilit\'e r\'eelle et complexe en m\'ecanique hamiltonienne"
N$^\circ$~JC05$_-$41465. .

%
\newcommand{\noopsort}[1]{}\def\polhk#1{\setbox0=\hbox{#1}{\ooalign{\hidewidth
  \lower1.5ex\hbox{`}\hidewidth\crcr\unhbox0}}} \def\cprime{$'$}
  \def\cydot{\leavevmode\raise.4ex\hbox{.}} \def\cprime{$'$} \def\cprime{$'$}
  \def\cprime{$'$} \def\cprime{$'$} \def\cprime{$'$}
  \def\polhk#1{\setbox0=\hbox{#1}{\ooalign{\hidewidth
  \lower1.5ex\hbox{`}\hidewidth\crcr\unhbox0}}} \def\cprime{$'$}
  \def\cprime{$'$} \def\cprime{$'$} \def\dbar{\leavevmode\hbox to
  0pt{\hskip.2ex \accent"16\hss}d}

\end{document}